\documentclass[%
prl
]{revtex4-1}
\usepackage{amsmath, amssymb, amstext}
\usepackage{graphicx}
\graphicspath{ {Plots/} }
\usepackage{ucs}
\usepackage[utf8x]{inputenc}
\usepackage[T1]{fontenc}
\usepackage[DIV=15]{typearea}
\usepackage{subcaption}
\usepackage{lettrine}
\usepackage{float}

\begin{document}


\begin{center}
{\bf\LARGE 
Conformer-specific polar cycloaddition of dibromobutadiene with trapped propene ions
}
\bigskip

Ardita Kilaj$^{1}$, Jia Wang$^{2\ast}$, Patrik Stra\v{n}\'ak$^{1\ast}$, Max Schwilk$^{3,1}$,  Ux\'{i}a Rivero$^{1}$, Lei Xu$^1$, \\ O. Anatole von Lilienfeld$^{3,1}$, Jochen K\"upper$^{2,4,5,6\dagger}$, and  Stefan Willitsch$^{1\dagger}$
\bigskip

$^1$ Department of Chemistry, University of Basel, Klingelbergstrasse 80, 4056 Basel, Switzerland \\
$^2$ Center for Free-Electron Laser Science, Deutsches Elektronen-Synchrotron DESY, Notkestrasse~85, 22607 Hamburg, Germany \\
$^3$ University of Vienna, Faculty of Physics, 1090 Vienna, Austria\\
$^4$ Department of Physics, Universit\"at Hamburg, Luruper Chaussee 149, 22761 Hamburg, Germany \\
$^5$ Department of Chemistry, Universit\"at Hamburg, Martin-Luther-King-Platz 6, 20146 Hamburg, Germany \\
$^6$ Center for Ultrafast Imaging, Universit\"at Hamburg, Luruper Chaussee 149,
22761 Hamburg, Germany \\
$\ast$ These authors contributed equally to the present work. \\
$\dagger$ Electronic mail: stefan.willitsch@unibas.ch,  jochen.kuepper@cfel.de
\end{center}


{\bfseries\noindent

Diels-Alder cycloadditions are efficient routes for the synthesis of cyclic organic compounds. There has been a long-standing discussion whether these reactions proceed via stepwise or concerted mechanisms. Here, we adopt a new experimental approach to explore the mechanistic details of the model polar cycloaddition of 2,3-dibromo-1,3-butadiene with propene ions by probing its conformational specificities in the entrance channel under single-collision conditions in the gas phase. Combining a conformationally controlled molecular beam with trapped ions, we find that both conformers of the diene, \emph{gauche} and \emph{s-trans}, are reactive with capture-limited reaction rates. Aided by quantum-chemical and quantum-capture calculations, this finding is rationalised by a simultaneous competition of concerted and stepwise reaction pathways, revealing an interesting mechanistic borderline case. 
}

\bigskip


It has been almost a century since Otto Diels and Kurt Alder described the cyclisation reaction which is now widely known as the Diels-Alder (DA) cycloaddition \cite{diels28a}. In this reaction, a conjugated diene and an alkene, the "dienophile", react to form a cyclohexene compound. Since its discovery in the 1920s, it has become one of the key reactions in synthetic organic chemistry for generating cyclic products \cite{ishihara14a}. Among others, it has been useful in the synthesis of many natural products such as the steroid hormone cortisone \cite{woodward52a}, reserpine, a drug for treating high blood pressure \cite{wender80a}, or the antibiotic tetracycline \cite{charest05a}.

The "canonical" mechanism of the DA cycloaddition assumes a concerted reaction proceeding via a single transition state in which bond formation and bond breaking occur synchronously \cite{houk95a, donoghue06a, black12a, sexton16a, rivero17a, rivero19a, rivero21a}. In this case, the transition state is presumed to be stabilised by H\"uckel aromaticity involving $[4+2]$ $\pi$ electrons which renders the concerted mechanism energetically more favourable than the other limiting scenario, a stepwise process with a diradical intermediate \cite{houk95a,donoghue06a}. The concerted mechanism explains why DA reactions often afford high stereo- and regioselectivity in the formation of the cycloadduct \cite{houk95a}. While often valid for symmetric systems, this picture breaks down in highly asymmetric systems in which the reaction becomes asynchronous to the extent that a stepwise mechanism is preferred \cite{domingo09a, linder12a}. In particular, this is the case for polar DA cycloadditions \cite{schmidt73a,bauld87a, eberlin04a} in which one of the reactants is charged. In the case of the $[4 + 1^{+}]$ polar cycloaddition, the removal of one electron from the dienophile leads to a radical cationic reaction in which the concerted transition state cannot be stabilised by H\"uckel aromaticity \cite{haberl99a,donoghue06a}. Owing to the practical importance of the DA reaction as well as its historical role in establishing quantum concepts in chemistry \cite{hoffmann68a}, its mechanism has been the subject of numerous experimental and theoretical investigations over the past decades and is still a subject of debate today \cite{sauer80a,houk95a,donoghue06a,domingo09a,black12a,linder12a,domingo16a}.

Experimental evidence for the mechanism of DA reactions has traditionally been obtained from kinetic isotope-substitution studies and from examinations of the regio- and stereospecificity of the reaction \cite{houk95a, singleton01a, bauld99a}. While the former method provides indirect evidence relying on the mechanistic interpretation of kinetic data, the latter only allows firm conclusions in the case of stepwise processes which are not stereoselective. A more direct and general route for the elucidation of the reaction mechanism can be provided by studies of the entrance channel of the reaction, in particular its conformational specificity. The concerted mechanism imposes that the reaction proceeds exclusively from the \emph{s-cis} (or possibly \emph{gauche}) conformer of the diene and not from the \emph{s-trans} conformer while a stepwise process would also enable the \emph{s-trans} species to participate in the formation of the cycloadduct. Thus, information about the details of the reaction mechanism could be obtained from measuring the reaction rates of the individual conformers of the diene. However, the experimental challenges in isolating individual molecular conformations and their tendency to interconvert under ambient conditions in solution have rendered this approach for probing the reaction mechanism elusive so far.

A promising route for the investigation of chemical reactions in a controlled environment has recently been established with experiments combining molecular beams \cite{chang15a, carrascosa17a} with trapped and Coulomb-crystallised molecular ions in the gas phase \cite{willitsch12a, heazlewood15a, willitsch17a}. Molecular beams generated by supersonic expansions allow molecular vibrations and internal rotations to be cooled down to very low temperatures such that individual molecular conformations are preserved. Their combination with inhomogeneous electrostatic fields has enabled the spatial separation of different conformers as well as individual rotational states based on their different electric dipole moments \cite{filsinger08a, filsinger09a, horke14a, trippel18a}. Directing such a controlled molecular beam at a reaction target of trapped ions has enabled kinetic and mechanistic studies of individual conformers of 3-aminophenol with Ca$^{+}$ ions \cite{chang13a, roesch14a} as well as nuclear-spin-selected water molecules with diazenylium ions ($\mathrm{N_2H^{+}}$) \cite{kilaj18a}. 

Here, we leveraged these methods for the investigation of the mechanism and kinetics of a prototypical polar cycloaddition reaction. We studied the $[4 + 1^{+}]$ cycloaddition of individual molecular conformations of 2,3-dibromobuta-1,3-diene (DBB) with propene ions ($\mathrm{C_3H_6^{+}}$) to form the 1,2-dibromo-4-methyl-cyclohexene radical cation (Fig.~\ref{fig_setup}a). DBB exists in an apolar \emph{s-trans} and a polar \emph{gauche} conformation (the \emph{s-cis} conformation constitutes a saddle point on the potential-energy surface (PES) of this molecule). We recently reported their successful electrostatic separation in a molecular beam \cite{kilaj20a}. In the conformer-specific reaction experiments presented here, we found that both the \emph{gauche} as well as the \emph{s-trans} conformer of DBB react efficiently with propene ions. 
Quantum-chemical calculations of the PES of the system reveal a highly asynchronous reaction with a simultaneous competition of stepwise and concerted reaction pathways involving both conformers. The present system can thus be understood as an interesting borderline case which simultaneously incorporates both limiting DA mechanisms, illustrating the mechanistic complexity of DA reactions in ionic systems. 


\section*{Results}

\subsection*{Experimental setup}

\begin{figure}[b!]
\centering
\includegraphics[width=100 mm]{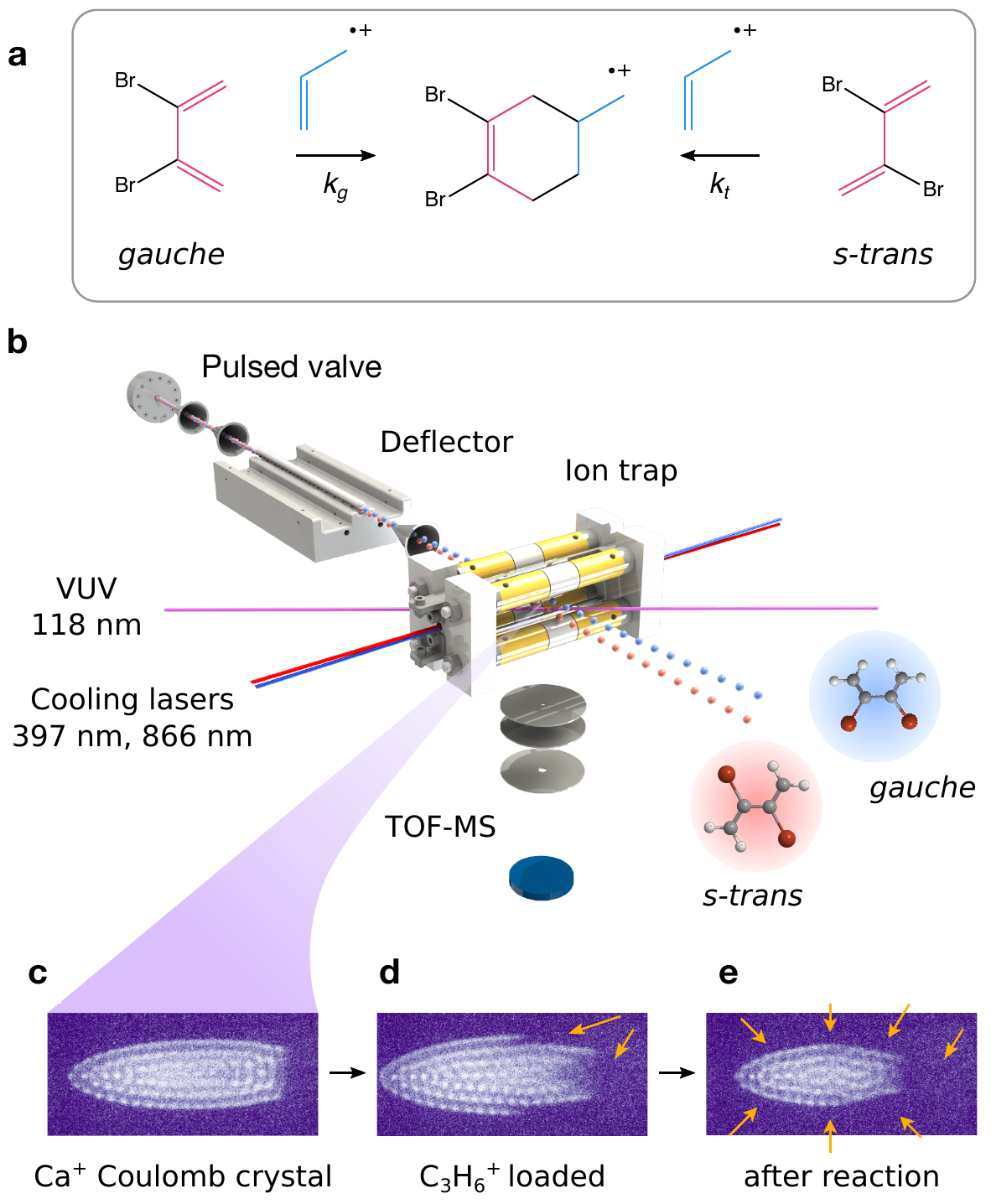}
\caption{\textbf{Overview of the experiment.} \textbf{a} Scheme of the polar cycloaddition reaction between the \emph{gauche} and \emph{s-trans} conformers of DBB with propene ions  exhibiting reaction rate constants $k_{g}$ and $k_{t}$, respectively. \textbf{b} Schematic of the experimental setup. The two conformers of DBB were separated by electrostatic deflection of a molecular beam and directed at an ion trap containing sympathetically cooled propene ions. Reaction products were measured by ion extraction into a time-of-flight mass spectrometer (TOF-MS). \textbf{c--e} Fluorescence images of laser-cooled Ca$^{+}$ Coulomb crystals at different stages of the experiment: \textbf{c} shows the initial, pure Ca$^{+}$ crystal, \textbf{d} shows an image after loading propene ions by vacuum-ultraviolet (VUV) photoionisation of propene, and \textbf{e} is a typical image obtained after the reaction with DBB. Arrows indicate regions where ions heavier than Ca$^{+}$ accumulate in the trap.}
\label{fig_setup}
\end{figure}

A schematic of the experimental setup is shown in Fig.~\ref{fig_setup}b. A cold beam of the neutral diene DBB in neon carrier gas was prepared by a pulsed supersonic expansion,  see Methods and \cite{kilaj20a}. The collimated beam passed through an electrostatic deflector in which a strong electric-field gradient induced a spatially varying Stark-energy shift for polar molecules \cite{chang15a, kilaj20a}. This resulted in a force acting on the polar \emph{gauche}-DBB (dipole moment $\mu_\mathrm{gauche} = 2.3$~D \cite{kilaj20a}) which was thus vertically deflected from the beam axis in contrast to the apolar \emph{s-trans} conformer ($\mu_\mathrm{trans} = 0$~D). 

The molecular beam with the spatially separated conformers was directed at a linear-quadrupole radiofrequency ion trap (LQT) which contained a Coulomb crystal of laser-cooled Ca$^{+}$ ions \cite{willitsch17a}. Fluorescence images of the Ca$^+$ ions were recorded using a charge-coupled device (CCD) camera, see Fig.~\ref{fig_setup}c. The Ca$^{+}$ Coulomb crystal served as a reservoir for the sympathetic cooling of propene ions as well as product ions formed during the reaction \cite{willitsch12a, willitsch17a}.

The Coulomb crystal formed a stationary reaction target for the molecular beam. By vertically tilting the molecular-beam assembly, different parts of the DBB beam were overlapped with the ions in the trap entailing reactions with samples of different compositions of the \emph{gauche} and \emph{s-trans} conformers of DBB. This enabled the study of the influence of the DBB conformation on the cycloaddition kinetics. The formation of reaction products and the decay of reactant ions was measured by ejecting the ions from the trap into a time-of-flight mass spectrometer (TOF-MS) after a defined reaction time.

\subsection*{Conformer separation of DBB}

\begin{figure}[t]
\centering
\includegraphics[width= \linewidth]{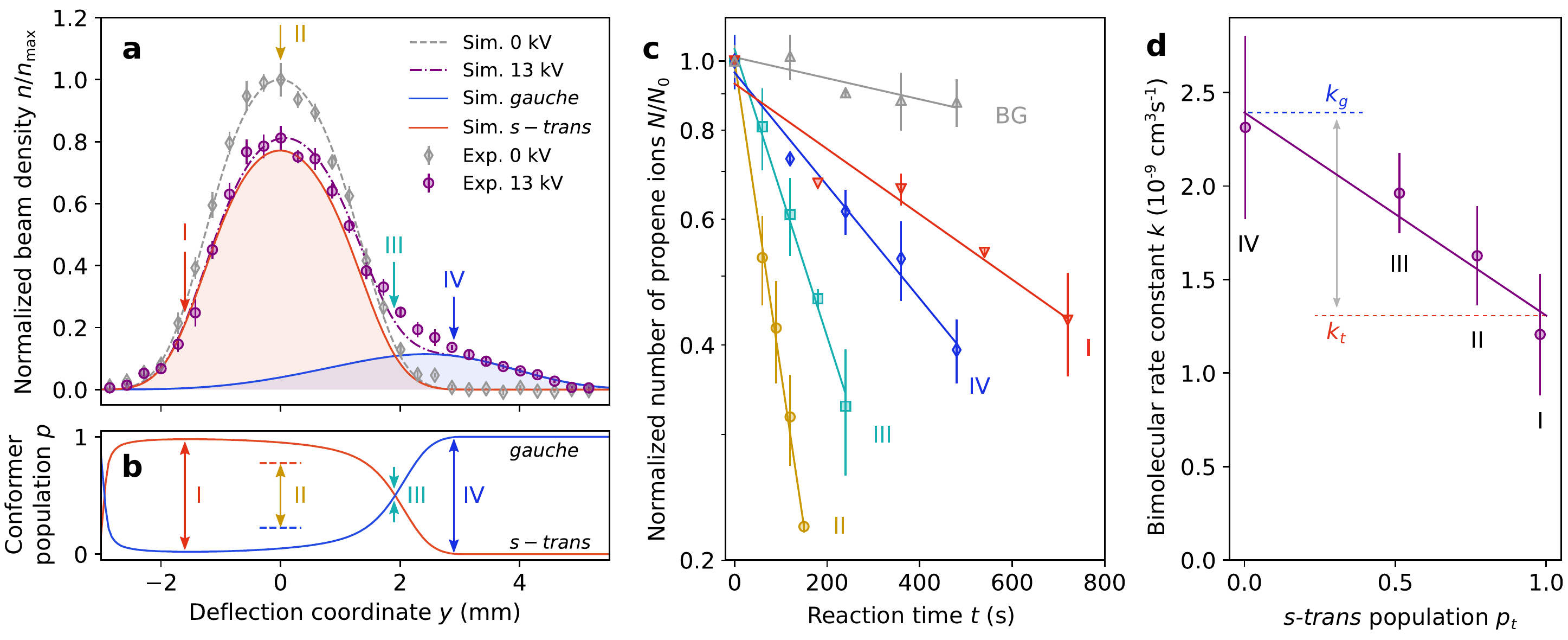}
\caption{\textbf{Conformer-specific reaction rate constants.} \textbf{a} Measurement of the DBB beam-density profile along the deflection coordinate for deflector voltages of 0~kV and 13~kV \cite{kilaj20a}. The experimental data (symbols) are in good agreement with Monte-Carlo trajectory simulations (lines). At a deflector voltage of 13~kV, the two conformers were partially separated in the molecular beam. \textbf{b} Populations of the conformers as a function of deflection coordinate in the molecular beam obtained from the trajectory simulations. \textbf{c} Reaction kinetics measured in terms of the normalised number of propene ions in the Coulomb crystal as a function of reaction time at the deflection coordinates I--IV marked in \textbf{a, b} and a background measurement (BG) without the molecular beam hitting the crystal. The exponential decay of the number of propene ions indicates a bimolecular reaction with pseudo-first-order kinetics. \textbf{d} Bimolecular rate constants extracted from the pseudo-first-order rate-constants as a function of the \emph{s-trans}-conformer population. The line represents a linear fit to the logarithmic data. Error bars correspond to one standard deviation of three independent measurements.}
\label{fig_rates}
\end{figure}

The density profile of the molecular beam of DBB was characterised along the vertical deflection axis. At specific tilt angles of the molecular beam, defining the deflection coordinate, the DBB molecules were ionised with vacuum-ultraviolet (VUV) radiation at a wavelength of 118~nm in the centre of the ion trap and subsequently ejected into the TOF-MS. 
Fig.~\ref{fig_rates}a shows normalised profiles of the beam density $n$ obtained as a function of the deflection coordinate $y$ measured with deflector voltages of 0~kV and 13~kV. The undeflected beam (grey symbols in Fig.~\ref{fig_rates}a) contained a 1:3.3 mixture of the \emph{gauche} and \emph{s-trans} conformers of DBB, respectively, determined by the thermal populations in the room temperature reservoir from which the molecular beam emanated \cite{kilaj20a}. 
Separating the polar \emph{gauche} from the apolar \emph{s-trans} conformer using the deflector led to the appearance of a shoulder in the density profile towards larger deflection coordinates (purple symbols in Fig.~\ref{fig_rates}a). The data are well reproduced by Monte-Carlo trajectory simulations of the molecular beam \cite{kilaj20a} (lines in Fig.~\ref{fig_rates}a) assuming a rotational temperature of 1~K and an independently measured beam velocity $v_\mathrm{beam} = 843(58)~\mathrm{m/s}$ \cite{kilaj20a}.
These simulations were used in order to determine the populations of the conformers as a function of the deflection coordinate (see Fig.~\ref{fig_rates}b) and to identify four positions, marked I--IV in Figs.~\ref{fig_rates}a and b, that corresponded to DBB samples with populations $p_{t}$ of the \emph{s-trans} conformer ranging from 1 to 0, respectively. Position II corresponds to the density maximum of the undeflected molecular beam (deflector voltage 0~kV) containing a thermal mixture of the conformers ($p_{t} = 0.77$). At positions I and IV, almost pure samples of \emph{s-trans} and \emph{gauche}-DBB were obtained. Position III corresponds to a mixture of practically equal contributions from both conformers.

\subsection*{Reaction rate measurements}

Exploiting the control over the conformational populations of DBB, reaction experiments were performed with the molecular beam at the deflection coordinates I--IV targeting the ion Coulomb crystal. The experimental sequence started by preparing a Coulomb crystal of about $10^3$ laser-cooled Ca$^{+}$ ions in the LQT. Subsequently, propene ions were loaded into the ion trap by leaking propene gas into the vacuum chamber at a partial pressure of $3\times10^{-9}$~mbar and photoionising propene molecules by VUV radiation. Due to their larger mass, the propene ions localised at the extremities of the Ca$^{+}$ ion crystal as evidenced by a change of its shape (Fig.~\ref{fig_setup}d). Note that no chemical reactions of neutral propene with the laser-cooled calcium ions were observed within the sensitivity limits of our experiment.

Reactions between propene ions and the DBB molecules were initiated by switching on the pulsed molecular beam. After a variable reaction time, the molecular beam was switched off. A typical Ca$^{+}$ fluorescence image taken after reaction (Fig.~\ref{fig_setup}e) shows a slight reduction of the number of Ca$^{+}$ ions and a spatial rearrangement of the crystal due to trapped product ions. The ions were ejected into the TOF-MS to determine the number $N$ of remaining propene ions. To discriminate between Ca$^{+}$ (mass 40 u) and C$_3$H$_6^+$ (mass 42 u), the TOF-MS was operated in a high-resolution mode within a limited mass range \cite{roesch16a}. 

Fig.~\ref{fig_rates}c shows measurements of relative C$_3$H$_6^+$ ion counts after reaction with DBB at the four molecular beam configurations I--IV as a function of reaction time. To account for the loss of propene ions due to collisions and reactions with background gas, background data sets were recorded for which the molecular beam was adjusted such that it did not impinge on the Coulomb crystal. An exemplary background measurement is shown by the grey data points in Fig.~\ref{fig_rates}c. Loss of propene ions due to collisions with neon from the molecular beam was found to be negligible in a control experiment with a pure beam of neon impacting on the crystal. All reaction measurements exhibited an exponential decay of the number of propene ions over time which implies a pseudo-first-order rate law for the bimolecular reaction, as expected for a constant DBB density replenished by the molecular beam. 

The pseudo-first-order rate constants $\tilde{k}_i$ at beam positions $i = \mathrm{I,\ldots,IV}$, were obtained from a linear fit to the logarithmic data (see Fig. \ref{fig_rates}c) and subtraction of the corresponding background rate constant. Bimolecular rate constants $k_i = \tilde{k}_i/n_i$ were calculated using the DBB beam densities $n_i$ determined from the beam density profile shown in Fig.~\ref{fig_rates}a and an independently measured absolute DBB density of $n_\mathrm{avg} = 3.9(4)\times 10^{6}~\mathrm{cm}^{-3}$ at $y = 0$ and deflector voltage 13~kV (see Supplementary Note S1).

\subsection*{Conformer-specific rate constants}

Fig.~\ref{fig_rates}d shows the measured bimolecular rate constants $k$ as a function of the \emph{s-trans} population $p_\mathrm{t}$ obtained from the Monte-Carlo trajectory simulations, Fig.~\ref{fig_rates}b. The bimolecular rate constant for the depletion of propene ions \emph{via} the two reactions in Fig.~\ref{fig_setup}a was modelled as a linear combination $k = p_{\mathrm{g},i} k_\mathrm{g} + p_{\mathrm{t},i} k_\mathrm{t}$ of the rate constants $k_\mathrm{g/t}$ of the individual \emph{gauche/s-trans}-conformers, respectively. The weighting factors $p_{\mathrm{g/t},i}$ correspond to the respective conformer populations at molecular beam position $i$. The solid line in Fig. \ref{fig_rates}d represents a least-squares fit using the linear model for $k$ which agrees with the experimental data within the error bars. The fit yields the bimolecular reaction rate constants $k_\mathrm{g} = 2.4(3)\times10^{-9}~\mathrm{cm^{3}s^{-1}}$ for the \emph{gauche} conformer and $k_\mathrm{t} = 1.3(2)\times10^{-9}~\mathrm{cm^{3}s^{-1}}$ for the \emph{s-trans} conformer. This implies that both conformers are reactive with \emph{gauche-}DBB reacting faster than \emph{s-trans}-DBB with a relative difference $r = 2(k_\mathrm{g} - k_\mathrm{t})/(k_\mathrm{g} + k_\mathrm{t}) = 0.6(1)$.

\subsection*{Reaction products}

\begin{figure}[!t]
\centering
\includegraphics[width=120 mm]{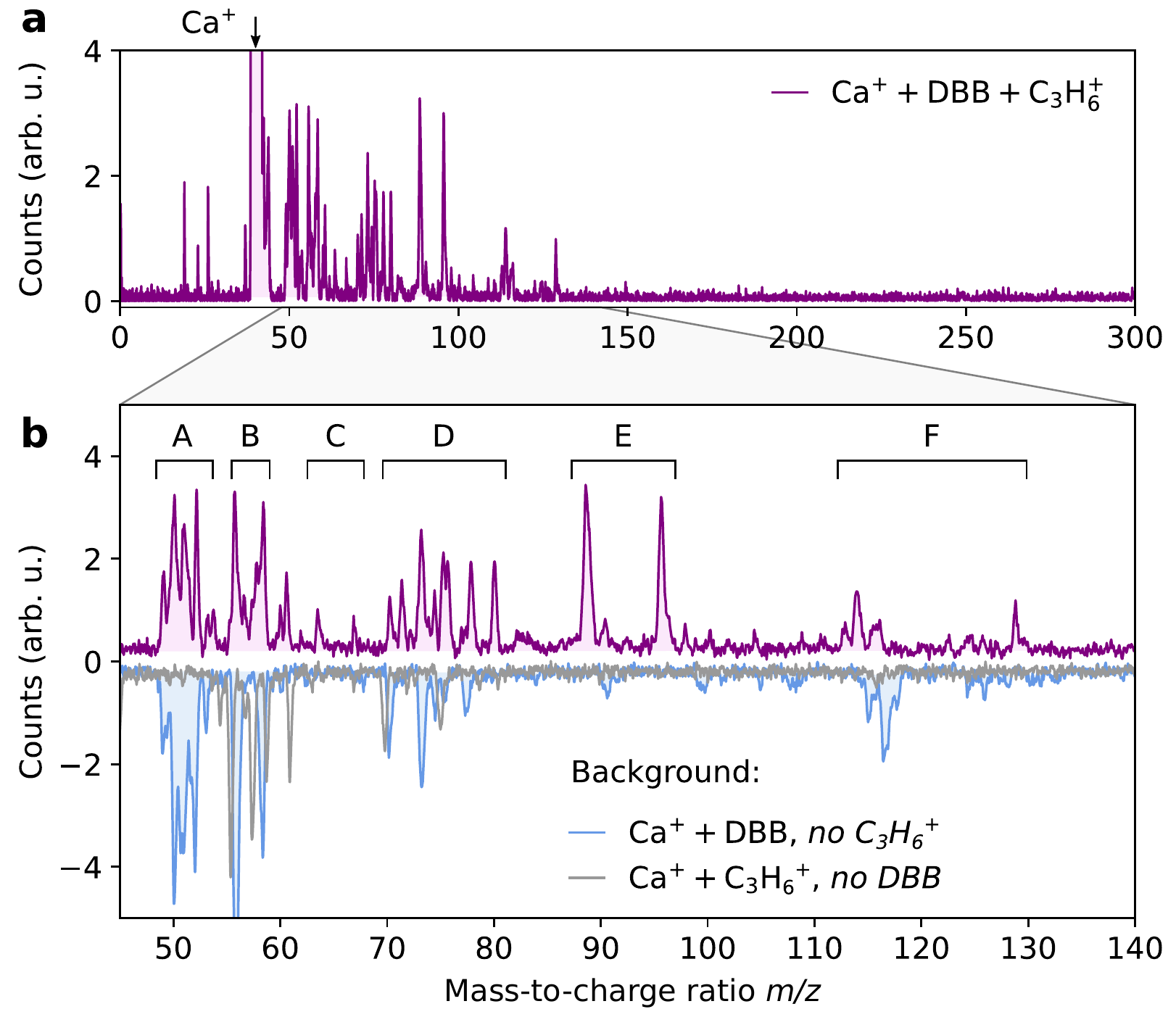}
\caption{\textbf{Mass spectra of reaction products.} \textbf{a} Overview time-of-flight mass spectrum acquired after 2~min of reaction of DBB with propene ions. The spectrum shows a strong signal of Ca$^{+}$ at 40~u and several product peaks. No signal could be detected at the expected mass of the cycloadduct at 252 and 256~u due to fragmentation. \textbf{b} Magnified view of the mass range showing the product signals (top trace) and comparison with control measurements in which either propene ions or DBB were absent from the experiment (lower inverted traces). The spectrum consists of several bands A--F corresponding to different fragments of the cycloadduct, see text for details.}
\label{fig_products_tofms}
\end{figure}

To gain information about the products of this reaction, mass spectra of the trapped ions were recorded after 2~min reaction time. These experiments were performed with the molecular beam set to deflection-coordinate position II where the high beam density enabled fast data acquisition. The full mass spectrum, averaged over 50 experiments, is shown in Fig.~\ref{fig_products_tofms}a. Due to a combination of the strong exothermicity of the reaction ($>60$~kcal/mol) and the constant presence of near-infrared and near-ultraviolet laser light used to cool Ca$^{+}$ (Methods), both of which promote fragmentation of the product ions, only fragments of the cycloadduct (mass $m=252, 256$~u) could be observed. This is consistent with the behaviour observed in radical cation reactions of similarly sized systems \cite{eberlin04a}. The dominant peak due to Ca$^{+}$ (40 u) was used to calibrate the mass scale of the spectra. An expanded view of the product-fragment spectrum is displayed in Fig.~\ref{fig_products_tofms}b and compared with control experiments either without propene ions in the trap (inverted blue trace) or without DBB in the molecular beam (inverted grey trace) under otherwise identical conditions. The TOF peaks can be grouped into several bands labelled with capital letters A--F (Supplementary Table S2). The bimodal structure of the peaks in the mass spectrum results from ions heavier than Ca$^{+}$ forming extended shells around the Ca$^{+}$ Coulomb crystal, so that they feel different extraction fields at different locations in the trap when accelerated into the TOF-MS. Therefore, the exact position and shape of the signals in the TOF-MS sensitively depends on the shapes and compositions of the multicomponent Coulomb crystals. The assignment of the signals in the TOF-MS to specific molecular compounds was based on detailed molecular dynamics (MD) simulations of the ejection process of the mixed species Coulomb crystals into the TOF-MS (see Supplementary Note S2). 

The first two bands A and B were attributed to $\mathrm{C_4H_n^{+}}$ ($m=50\text{-}52$~u) and CaO$^+$ ($m=56$~u), $\mathrm{CaOH^{+}}$ ($m=57$~u), respectively, which were also present with the same intensities in the control experiments. This shows that $\mathrm{C_4H_n^{+}}$ fragments were mainly formed by reactive collisions between Ca$^{+}$ and DBB. Similarly, the band F was assigned to CaBr$^{+}$ ($m=119$~u and 121~u) which also appeared in the control experiment in which C$_3$H$_6^{+}$ was absent. CaO$^+$ and CaOH$^+$ were likely formed in a reaction of the Ca$^+$ ions with residual H$_2$O background gas in the vacuum chamber.

Fragments of the reaction product of DBB with the propene ions appeared as bands C, D and E.
The strongest signal is observed for band E and can be attributed to a $\mathrm{C_7H_9^{+}}$ ($m=93$~u) fragment of the cycloadduct $\mathrm{C_7H_{10}Br_2^{+}}$, after removal of the two bromines and a hydrogen. Band D has a more complicated structure and is likely explained by the presence of $\mathrm{C_6H_5^{+}}$ ($m=77$~u) and $\mathrm{C_6H_6^{+}}$ ($m=78$~u). Finally, band C is consistent with the fragment $\mathrm{C_5H_5^{+}}$ ($m=65$~u) which may be formed by loss of a formal $\mathrm{C_2H_4}$ unit from the $\mathrm{C_7H_{9}^{+}}$ ($m=93$~u) intermediate. The gradual decrease of signal intensity with decreasing mass from bands E to C suggests a stepwise fragmentation of the parent product ion. A more detailed discussion of the possible fragmentation pathways is given in Supplementary Note S2. In summary, the observation of all of these fragments is strong evidence that a DA cycloadduct is indeed formed in the reaction of DBB with propene ions.


\section*{Discussion}

Our experiments revealed a strong reactivity of propene ions with both conformers of DBB. Moreover, a pronounced enhancement of the reaction rate was observed with DBB in its \emph{gauche} conformation. In order to understand the origin of this reactivity, quantum chemical calculations of the PES of this reaction were performed using density functional theory (Methods). Since propene has a larger ionization potential than DBB, the asymptote DBB + C$_3$H$_6^+$ accessed in the experiments corresponds to the first electronically excited state of the ionic reaction system $\approx 8$~kcal/mol above the ground-state asymptote. However, through long-range charge exchange via a conical intersection with the ground state early in the entrance channel, the reaction subsequently proceeds on the ground-state surface which asymptotically connects to $\rm DBB^{+}$ + C$_3$H$_6$.

\begin{figure}[t!]
\centering
\includegraphics[width=\textwidth]{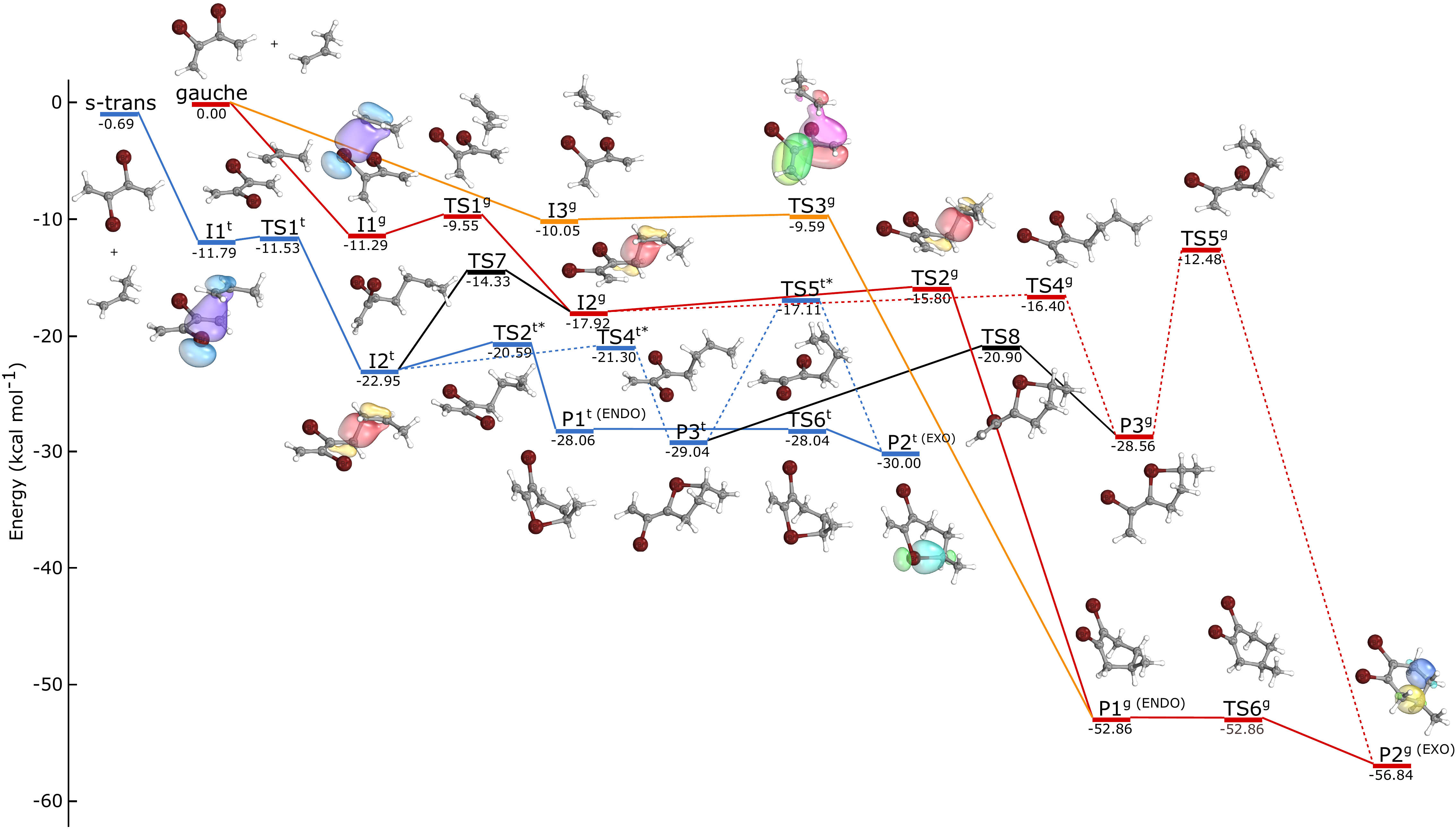}
\caption{\textbf{Zero-point-corrected potential-energy profile} of the electronic ground state of the [DBB-C$_3$H$_6]^+$ system showing minimum-energy paths for the polar cycloadditions of both conformers of DBB calculated using the M06-2X density functional and the def2-TZVPP basis set (Methods). Structures marked with an asterisk have been identified in additional conformations which are not displayed for clarity. Selected molecular orbitals which illustrate chemical bonds formed during the reaction  are depicted for specific structures.}
\label{fig_theory}
\end{figure}

Fig.~\ref{fig_theory} shows zero-point-corrected energies of stationary points on the ground-state PES for reaction pathways of either conformer. For specific structures, selected molecular orbitals are depicted which illustrate relevant chemical bonds formed during the reaction. For \emph{gauche}-DBB, two pathways connecting to the \emph{endo} and \emph{exo} conformations of the cycloaddition product were found. One of these pathways proceeds via a single transition state (TS3$^\text{g}$) and can therefore be classified as concerted \cite{donoghue06a}. However, this pathway is strongly asynchronous as TS3$^\text{g}$ was found to be a loose transition state with a highly asymmetric structure exhibiting considerably different lengths of the newly formed C-C bonds (3.03~\AA~vs. 4.86~\AA). By contrast, the second pathway is stepwise involving two separate bond-formation steps via the transition states TS1$^\text{g}$ and TS2$^\text{g}$ connected by intermediate I2$^\text{g}$. Moreover, from I2$^\text{g}$ a five-membered-ring structure P3$^\text{g}$ can be formed which contains a hypervalent bromine atom and can rearrange to the DA adduct P2$^\text{g}$ via TS5$^\text{g}$.

For the \emph{s-trans} conformer of DBB, a stepwise pathway connecting to the DA cycloadducts via intermediate I2$^\text{t}$ was identified. From I2$^\text{t}$, the \emph{s-trans} path folds into the \emph{gauche} path through a conformational rearrangement connecting to I2$^\text{g}$ via TS7. Additionally, the isomeric five- and six-membered-ring structures P1$^\text{t}$, P2$^\text{t}$ and P3$^\text{t}$ containing hypervalent bromine atoms are also accessible starting from I2$^\text{t}$. These, however, are energetically not stable and will eventually isomerise to the DA adducts or fragment. Considering the substantial total energy release of more than 60~kcal/mol in the reaction, it can be expected that also the DA adduct will fragment under the present conditions, as was indeed observed experimentally. 

Based on these theoretical calculations, it can be concluded that the polar cycloaddition of both conformers is energetically feasible in the present experiments, in line with the experimental findings. All barriers identified along the minimum-energy paths are submerged, i.e., their heights are below the energy of the reactants, thus rendering all identified pathways effectively barrierless. 

\begin{figure}[t!]
\centering
\includegraphics[width= 130 mm]{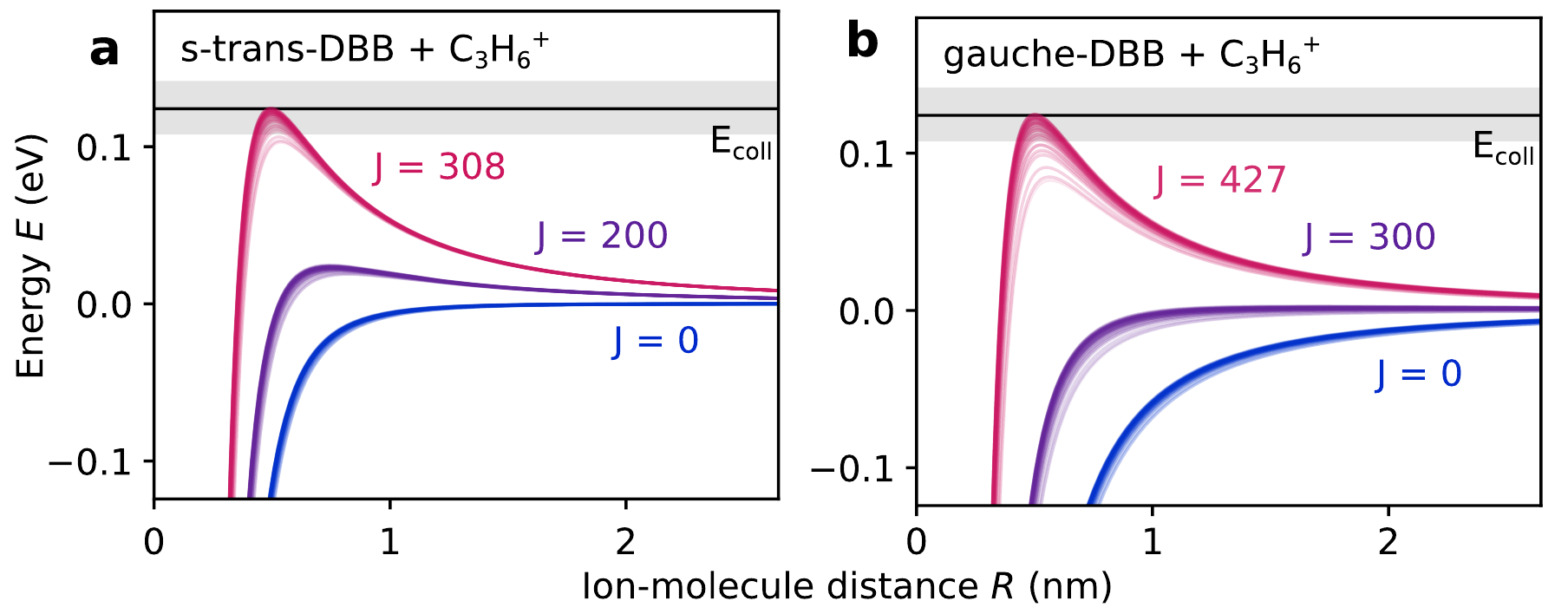}
\caption{\textbf{Rotationally adiabatic, centrifugally corrected ion-molecule interaction potentials} for the reaction of \textbf{a} \emph{s-trans}- and \textbf{b} \emph{gauche}-DBB with propene ions for different values of the total collisional angular momentum quantum number $J$. Each value of $J$ comprises a set of lines corresponding to all rotational states of DBB with rotational angular momentum quantum number $j=4$ which corresponds to the largest population in the molecular beam. The collision energy $E_\mathrm{coll}$ is indicated by the black horizontal line and the grey-shaded areas represent its experimental uncertainty.}
\label{fig_theory2}
\end{figure}

Motivated by these findings, the reaction kinetics were modelled using rotationally adiabatic quantum capture theory for barrierless ion-molecule reactions \cite{clary87a, stoecklin92a}. Within this framework, it is assumed that the short-range reaction occurs with unit probability and that the overall reaction rate is governed by the long-range interactions between the ionic and neutral collision partners and the centrifugal barrier. Figs.~\ref{fig_theory2}a and b show rotationally adiabatic, centrifugally corrected long-range interaction potentials for collisions of C$_3$H$_6^+$ with \emph{gauche}- and \emph{s-trans}-DBB, respectively. They include the interaction of the ionic charge to the induced and permanent dipole moments of the neutral molecules. The three sets of curves correspond to different values of the total collisional angular momentum quantum number $J$ which gives rise to the centrifugal energy barrier. The individual curves for each value of $J$ correspond to all rotational quantum states of DBB with rotational angular momentum quantum number $j=4$ which was populated the most in the molecular beam.

A unit reaction probability was assigned to any collision with $J<J_\mathrm{max}$ \cite{clary87a,stoecklin92a}, where $J_\mathrm{max}$ is the maximum angular momentum quantum number for which the centrifugally corrected interaction energy does not exceed the experimental collision energy $E_\mathrm{coll} = 124(17)$~meV (black solid line in Figs.~\ref{fig_theory2}a,b). The plots show that the centrifugal barrier grows faster with $J$ for \emph{s-trans}-DBB than for \emph{gauche}-DBB such that the maximum collisional angular momenta for a reactive encounter are $J_\mathrm{max, t} \approx 308$ and $J_\mathrm{max, g} \approx 427$, respectively, implying a larger cross-section proportional to $J_\mathrm{max}^2$ \cite{stoecklin92a} for \emph{gauche}-DBB. This is due to the attractive interaction of the ion with the dipole moment of \emph{gauche}-DBB, which is not present for the apolar \emph{s-trans} conformer. As a result, adiabatic capture theory predicts bimolecular rate constants of $k_\mathrm{ac,g} = 2.4 \times10^{-9}~\mathrm{cm^{3}s^{-1}}$ for \emph{gauche}-DBB and $k_\mathrm{ac,t} = 1.3 \times10^{-9}~\mathrm{cm^{3}s^{-1}}$ for \emph{s-trans}-DBB, in very good agreement with the measured values. 
The calculated relative difference of the reaction rate constants between the \emph{gauche} and \emph{s-trans} conformers is $r_\mathrm{ac} = 0.64 $ which also agrees well with the experimental value $r = 0.6(1)$. The good agreement between the theoretical and experimental rate constants vindicates the mechanistic picture drawn from the quantum-chemical calculations above.


In conclusion, we have presented results on the chemical reactivity of individual molecular conformations in the prototypical polar cycloaddition reaction of 2,3-dibromobutadiene with propene ions. The current study represents the first application of conformationally controlled molecular beams and ion-trapping methods for studies of mechanistic details of organic name reactions under single-collision conditions in the gas phase. Our results indicate that both the \emph{gauche} as well as the \emph{s-trans} conformers of DBB react efficiently. This result is at variance with a purely concerted mechanism for the reaction in which only the \emph{gauche} conformer is expected to react. The observed reactivity of both molecular conformations of the diene in conjunction with quantum-chemical calculations indicates the presence of stepwise reaction pathways for the cycloaddition in competition with a concerted process. We thus conclude that both limiting scenarios of the DA reaction mechanism are simultaneously active in the present system. The almost twofold larger rate constant for the \emph{gauche} compared to the \emph{s-trans} species was explained by attractive long-range ion-molecule interactions in very good quantitative agreement with an adiabatic capture model. 

The present work demonstrated a platform for probing chemical kinetics with individual molecular conformations in models systems of fundamental organic name reactions under controlled conditions in the gas phase. In combination with mass-spectrometric tools, this capability opens up a wide range of opportunities for the precise investigation and validation of reaction mechanisms.


\section*{Methods}
\label{sec:methods}

\subsection*{Molecular beam}
The molecular beam was generated from DBB vapour at room temperature seeded in neon carrier gas at 5~bar. The gas mixture was pulsed through a cantilever piezo valve (MassSpecpecD ACPV2, $150~\mu$m nozzle diameter) at a repetition rate of 200~Hz with a gas-pulse duration of $250~\mu$s measured at the position of the LQT. The speed of the resulting molecular beam was measured to be $v_\mathrm{beam} = 843(58)~\mathrm{m/s}$ \cite{kilaj20a}. 

\subsection*{Ion trap and TOF-MS}
The LQT was operated at a peak-to-peak radio frequency (RF) voltage of $V_\mathrm{RF,pp} = 800$~V and a frequency $\Omega_\mathrm{RF} = 2\pi \times 3.304~\mathrm{MHz}$. Laser cooling of Ca$^+$ ions was achieved using two continuous-wave laser beams at 397~nm and 866~nm generated by frequency-stabilised external-cavity diode lasers \cite{willitsch12a}. The Coulomb crystals were imaged by collecting the spatially resolved cooling-laser-induced fluorescence of the Ca$^{+}$ ions with a microscope onto a camera (Fig.~\ref{fig_setup}c--e). 

The LQT was coupled to a TOF-MS orthogonal to the molecular-beam propagation axis for the mass and quantitative analysis of reactant and product ions \cite{roesch16a}. 
For the measurement of overview mass spectra of the reaction products, a low-resolution operation mode was used for the extraction of ions into the TOF-MS by applying a $1~\mu$s long voltage pulse with a magnitude of 4.0~kV to the repeller electrodes. For the reaction-rate measurements, the resolution was enhanced by applying an additional pulse of 4.0~kV, delayed by $0.45~\mu$s, to the extractor electrodes \cite{roesch16a}. Ions were detected by a microchannel plate (MCP, Photonis USA) detector operating at a typical voltage of 2.3 kV placed at the end of the flight tube.

\subsection*{Femtosecond-laser and VUV ionisation}

Ionisation of DBB molecules and Ca atoms was performed with pulses from a Ti:Sapphire femtosecond laser (CPA 2110, Clark-MXR, Inc.) at a wavelength of 775~nm and a pulse duration of 150~fs focused at the centre of the LQT. Ca atoms were ionised before loading into the ion trap in a standardised procedure. A constant size and composition of the crystals was verified by TOF-MS. Similarly, DBB from the molecular beam was ionised using femtosecond laser pulses to determine the beam density (see Supplementary Note S1). In addition, a pulsed vacuum-ultraviolet (VUV) light source at 118 nm, focused down to a spot size of $\approx 100~\mu$m, was used for soft ionisation of the propene molecules and to measure the DBB deflection profile \cite{kilaj20a}.

\subsection*{Quantum-chemical calculations}
Quantum-chemical calculations of the potential energy surface were performed by spin unrestricted density functional theory (DFT) using the M06-2X functional \cite{zhao08b}, which was evaluated for the treatment of DA reactions in Refs. \cite{linder13a, rivero17a}, with the 
def2-TZVPP basis set \cite{weigend05a} using the Gaussian 09 software package \cite{g09}. Stuttgart effective core potentials (ECP) including scalar relativistic effects were used for 10 inner-shell electrons of the bromine atoms on DBB \cite{bergner1993ab}. Additionally, the main stationary points were also found to be consistent with corresponding calculations based on the B3LYP functional \cite{lee88b,becke93a} with the exception of P1$^{g}$ and P1$^{t}$ which were not found to be local minima on the B3LYP level and directly converged to the P2$^{g}$ and P2$^{t}$ structures, respectively. Moreover, the present DFT methodology was extensively validated against coupled-cluster and multireference methods with different basis sets in the related DBB + Ca$^+$ reaction system, the details of which will be published elsewhere. Intrinsic reaction coordinates were computed to connect all transition states with intermediates shown in Fig. \ref{fig_theory}. The similar energy of TS6$^{g}$ and P1$^{g}$ arises from the addition of the zero-point energy (ZPE) and the very shallow nature of the TS. The single-point energies of the structures are listed in Supplementary Table~S3. Bonding and wavefunction analyses were performed using intrinsic bond orbitals at the M06-2X/def2-TZVPP level with the IboView program package \cite{knizia2013}. The spin contamination of the wavefunctions of the stationary points has been found to be small (see Supplementary Table ~S3), validating the application of the present single-reference level of theory.


\section*{Acknowledgements}
We thank Philipp Kn\"opfel, Grischa Martin, Georg Holderried and Anatoly Johnson (University of Basel) for technical support. This work is supported by the Swiss National Science Foundation under grants no. BSCGI0\_157874 and IZCOZ0\_189907. We acknowledge support by the University of Basel and by Deutsches Elektronen-Synchrotron DESY, a member of the Helmholtz Association (HGF).  J.W.\ acknowledges a fellowship in the Helmholtz-OCPC postdoctoral exchange program. O.A.v.L. and M.S. acknowledge funding received from the European Union's Horizon 2020 research and innovation programme under Grant Agreements \#957189 and \#772834.

\section*{Author contributions}
S.W.\ and J.K.\ conceived the project, A.K.\ and J.W.\ performed the experiments. A.K.\ analysed the data, performed capture-rate calculations, and simulated molecular-beam profiles. P.S., M.S. and U.R. performed the quantum-chemistry calculations. L.X.\ performed the simulations of the TOF mass spectra. A.K. and S.W. drafted the manuscript. S.W., J.K.\ and O.A.v.L.\ supervised the project. All authors contributed to and approved the final manuscript.



\bibliography{all_refs}

\begin{thebibliography}{10}
\expandafter\ifx\csname url\endcsname\relax
  \def\url#1{\texttt{#1}}\fi
\expandafter\ifx\csname urlprefix\endcsname\relax\def\urlprefix{URL }\fi
\providecommand{\bibinfo}[2]{#2}
\providecommand{\eprint}[2][]{\url{#2}}

\bibitem{diels28a}
\bibinfo{author}{Diels, O.} \& \bibinfo{author}{Alder, K.}
\newblock \bibinfo{title}{Synthesen in der hydroaromatischen {R}eihe}.
\newblock \emph{\bibinfo{journal}{Justus Liebigs Ann. Chem.}}
  \textbf{\bibinfo{volume}{460}}, \bibinfo{pages}{98--122}
  (\bibinfo{year}{1928}).

\bibitem{ishihara14a}
\bibinfo{author}{Ishihara, K.} \& \bibinfo{author}{Sakakura, A.}
\newblock \emph{\bibinfo{title}{{Comprehensive Organic Synthesis}}},
  vol.~\bibinfo{volume}{5} (\bibinfo{publisher}{Elsevier},
  \bibinfo{address}{Oxford}, \bibinfo{year}{2014}), \bibinfo{edition}{2} edn.

\bibitem{woodward52a}
\bibinfo{author}{Woodward, R.~B.}, \bibinfo{author}{Sondheimer, F.},
  \bibinfo{author}{Taub, D.}, \bibinfo{author}{Heusler, K.} \&
  \bibinfo{author}{McLamore, W.~M.}
\newblock \bibinfo{title}{The total synthesis of steroids}.
\newblock \emph{\bibinfo{journal}{J. Am. Chem. Soc.}}
  \textbf{\bibinfo{volume}{74}}, \bibinfo{pages}{4223--4251}
  (\bibinfo{year}{1952}).

\bibitem{wender80a}
\bibinfo{author}{Wender, P.~A.}, \bibinfo{author}{Schaus, J.~M.} \&
  \bibinfo{author}{White, A.~W.}
\newblock \bibinfo{title}{General methodology for cis-hydroisoquinoline
  synthesis: synthesis of reserpine}.
\newblock \emph{\bibinfo{journal}{J. Am. Chem. Soc.}}
  \textbf{\bibinfo{volume}{102}}, \bibinfo{pages}{6157--6159}
  (\bibinfo{year}{1980}).

\bibitem{charest05a}
\bibinfo{author}{Charest, M.~G.}, \bibinfo{author}{Siegel, D.~R.} \&
  \bibinfo{author}{Myers, A.~G.}
\newblock \bibinfo{title}{Synthesis of (-)-tetracycline}.
\newblock \emph{\bibinfo{journal}{J. Am. Chem. Soc.}}
  \textbf{\bibinfo{volume}{127}}, \bibinfo{pages}{8292--8293}
  (\bibinfo{year}{2005}).

\bibitem{houk95a}
\bibinfo{author}{Houk, K.~N.}, \bibinfo{author}{Gonz\'alez, J.} \&
  \bibinfo{author}{Li, Y.}
\newblock \bibinfo{title}{Pericyclic reaction transition states: Passions and
  punctilios, 1935-1995}.
\newblock \emph{\bibinfo{journal}{Acc. Chem. Res.}}
  \textbf{\bibinfo{volume}{28}}, \bibinfo{pages}{81} (\bibinfo{year}{1995}).

\bibitem{donoghue06a}
\bibinfo{author}{Donoghue, P.~J.} \& \bibinfo{author}{Wiest, O.}
\newblock \bibinfo{title}{Structure and reactivity of radical ions: New twists
  on old concepts}.
\newblock \emph{\bibinfo{journal}{Chem. Eur. J.}}
  \textbf{\bibinfo{volume}{12}}, \bibinfo{pages}{7018--7026}
  (\bibinfo{year}{2006}).

\bibitem{black12a}
\bibinfo{author}{Black, K.}, \bibinfo{author}{Liu, P.}, \bibinfo{author}{Xu,
  L.}, \bibinfo{author}{Doubleday, C.} \& \bibinfo{author}{Houk, K.~N.}
\newblock \bibinfo{title}{Dynamics, transition states, and timing of bond
  formation in {D}iels–{A}lder reactions}.
\newblock \emph{\bibinfo{journal}{Proc. Natl. Acad. Sci. U.S.A.}}
  \textbf{\bibinfo{volume}{109}}, \bibinfo{pages}{12860}
  (\bibinfo{year}{2012}).

\bibitem{sexton16a}
\bibinfo{author}{Sexton, T.}, \bibinfo{author}{Kraka, E.} \&
  \bibinfo{author}{Cremer, D.}
\newblock \bibinfo{title}{Extraordinary mechanism of the {D}iels−{A}lder
  reaction: Investigation of stereochemistry, charge transfer, charge
  polarization, and biradicaloid formation}.
\newblock \emph{\bibinfo{journal}{J. Phys. Chem. A}}
  \textbf{\bibinfo{volume}{120}}, \bibinfo{pages}{1097} (\bibinfo{year}{2016}).

\bibitem{rivero17a}
\bibinfo{author}{Rivero, U.}, \bibinfo{author}{Meuwly, M.} \&
  \bibinfo{author}{Willitsch, S.}
\newblock \bibinfo{title}{A computational study of the {D}iels-{A}lder
  reactions between 2, 3-dibromo-1, 3-butadiene and maleic anhydride}.
\newblock \emph{\bibinfo{journal}{Chem. Phys. Let.}}
  \textbf{\bibinfo{volume}{683}}, \bibinfo{pages}{598--605}
  (\bibinfo{year}{2017}).

\bibitem{rivero19a}
\bibinfo{author}{Rivero, U.}, \bibinfo{author}{Unke, O.~T.},
  \bibinfo{author}{Meuwly, M.} \& \bibinfo{author}{Willitsch, S.}
\newblock \bibinfo{title}{Reactive atomistic simulations of {D}iels-{A}lder
  reactions: The importance of molecular rotations}.
\newblock \emph{\bibinfo{journal}{J. Chem. Phys.}}
  \textbf{\bibinfo{volume}{151}}, \bibinfo{pages}{104301}
  (\bibinfo{year}{2019}).

\bibitem{rivero21a}
\bibinfo{author}{Rivero, U.}, \bibinfo{author}{Turan, H.~T.},
  \bibinfo{author}{Meuwly, M.} \& \bibinfo{author}{Willitsch, S.}
\newblock \bibinfo{title}{Reactive atomistic simulations of
  {D}iels-{A}lder-type reactions: Conformational and dynamic effects in the
  polar cycloaddition of 2,3-dibromobutadiene radical ions with maleic
  anhydride}.
\newblock \emph{\bibinfo{journal}{Mol. Phys.}} \textbf{\bibinfo{volume}{119}},
  \bibinfo{pages}{e1825852} (\bibinfo{year}{2021}).

\bibitem{domingo09a}
\bibinfo{author}{Domingo, L.~R.} \& \bibinfo{author}{S\'aez, J.~A.}
\newblock \bibinfo{title}{Understanding the mechanism of polar {D}iels-{A}lder
  reactions}.
\newblock \emph{\bibinfo{journal}{Org. Biomol. Chem.}}
  \textbf{\bibinfo{volume}{7}}, \bibinfo{pages}{3576} (\bibinfo{year}{2009}).

\bibitem{linder12a}
\bibinfo{author}{Linder, M.} \& \bibinfo{author}{Brinck, T.}
\newblock \bibinfo{title}{Stepwise {D}iels−{A}lder: More than just an oddity?
  a computational mechanistic study}.
\newblock \emph{\bibinfo{journal}{J. Org. Chem.}}
  \textbf{\bibinfo{volume}{77}}, \bibinfo{pages}{6563} (\bibinfo{year}{2012}).

\bibitem{schmidt73a}
\bibinfo{author}{Schmidt, R.~R.}
\newblock \bibinfo{title}{Polar cycloadditions}.
\newblock \emph{\bibinfo{journal}{Angew. Chem. Int. Ed. Engl.}}
  \textbf{\bibinfo{volume}{12}}, \bibinfo{pages}{212--224}
  (\bibinfo{year}{1973}).

\bibitem{bauld87a}
\bibinfo{author}{Bauld, N.~L.} \emph{et~al.}
\newblock \bibinfo{title}{Cation radical pericyclic reactions}.
\newblock \emph{\bibinfo{journal}{Acc. Chem. Res.}}
  \textbf{\bibinfo{volume}{20}}, \bibinfo{pages}{371} (\bibinfo{year}{1987}).

\bibitem{eberlin04a}
\bibinfo{author}{Eberlin, M.~N.}
\newblock \bibinfo{title}{Gas-phase polar cycloadditions}.
\newblock \emph{\bibinfo{journal}{Int. J. Mass Spectrom.}}
  \textbf{\bibinfo{volume}{235}}, \bibinfo{pages}{263--278}
  (\bibinfo{year}{2004}).

\bibitem{haberl99a}
\bibinfo{author}{Haberl, U.}, \bibinfo{author}{Wiest, O.} \&
  \bibinfo{author}{Steckhan, E.}
\newblock \bibinfo{title}{Ab initio studies of the radical cation
  {D}iels-{A}lder reaction}.
\newblock \emph{\bibinfo{journal}{J. Am. Chem. Soc.}}
  \textbf{\bibinfo{volume}{121}}, \bibinfo{pages}{6730--6736}
  (\bibinfo{year}{1999}).

\bibitem{hoffmann68a}
\bibinfo{author}{Hoffmann, R.} \& \bibinfo{author}{Woodward, R.~B.}
\newblock \bibinfo{title}{The conservation of orbital symmetry}.
\newblock \emph{\bibinfo{journal}{Acc. Chem. Res.}}
  \textbf{\bibinfo{volume}{1}}, \bibinfo{pages}{17} (\bibinfo{year}{1968}).

\bibitem{sauer80a}
\bibinfo{author}{Sauer, J.} \& \bibinfo{author}{Sustmann, R.}
\newblock \bibinfo{title}{Mechanistic aspects of {D}iels-{A}lder reactions: A
  critical survey}.
\newblock \emph{\bibinfo{journal}{Angew. Chem. Int. Ed. Engl.}}
  \textbf{\bibinfo{volume}{19}}, \bibinfo{pages}{779--807}
  (\bibinfo{year}{1980}).

\bibitem{domingo16a}
\bibinfo{author}{Domingo, L.~R.}, \bibinfo{author}{R\'{i}os-Guti\'{e}rrez, M.},
  \bibinfo{author}{Chamorro, E.} \& \bibinfo{author}{P\'{e}rez, P.}
\newblock \bibinfo{title}{Aromaticity in pericyclic transition state
  structures? {A} critical rationalisation based on the topological analysis of
  electron density}.
\newblock \emph{\bibinfo{journal}{ChemistrySelect}}
  \textbf{\bibinfo{volume}{1}}, \bibinfo{pages}{6026} (\bibinfo{year}{2016}).

\bibitem{singleton01a}
\bibinfo{author}{Singleton, D.~A.} \emph{et~al.}
\newblock \bibinfo{title}{Isotope effects and the distinction between
  synchronous, asynchronous and stepwise {D}iels-{A}lder reactions}.
\newblock \emph{\bibinfo{journal}{Tethraedron}} \textbf{\bibinfo{volume}{57}},
  \bibinfo{pages}{5149} (\bibinfo{year}{2001}).

\bibitem{bauld99a}
\bibinfo{author}{Bauld, N.} \& \bibinfo{author}{Yang, J.}
\newblock \bibinfo{title}{A two step, non-stereospecific cation radical
  {D}iels-{A}lder reaction}.
\newblock \emph{\bibinfo{journal}{Tet. Lett.}} \textbf{\bibinfo{volume}{40}},
  \bibinfo{pages}{8519} (\bibinfo{year}{1999}).

\bibitem{chang15a}
\bibinfo{author}{Chang, Y.-P.}, \bibinfo{author}{Horke, D.~A.},
  \bibinfo{author}{Trippel, S.} \& \bibinfo{author}{K{\"u}pper, J.}
\newblock \bibinfo{title}{Spatially-controlled complex molecules and their
  applications}.
\newblock \emph{\bibinfo{journal}{Int. Rev. Phys. Chem.}}
  \textbf{\bibinfo{volume}{34}}, \bibinfo{pages}{557} (\bibinfo{year}{2015}).

\bibitem{carrascosa17a}
\bibinfo{author}{Carrascosa, E.}, \bibinfo{author}{Meyer, J.} \&
  \bibinfo{author}{Wester, R.}
\newblock \bibinfo{title}{Imaging the dynamics of ion-molecule reactions}.
\newblock \emph{\bibinfo{journal}{Chem. Soc. Rev}}
  \textbf{\bibinfo{volume}{46}}, \bibinfo{pages}{7498--7516}
  (\bibinfo{year}{2017}).

\bibitem{willitsch12a}
\bibinfo{author}{Willitsch, S.}
\newblock \bibinfo{title}{Coulomb-crystallised molecular ions in traps:
  methods, applications, prospects}.
\newblock \emph{\bibinfo{journal}{Int. Rev. Phys. Chem.}}
  \textbf{\bibinfo{volume}{31}}, \bibinfo{pages}{175--199}
  (\bibinfo{year}{2012}).

\bibitem{heazlewood15a}
\bibinfo{author}{Heazlewood, B.} \& \bibinfo{author}{Softley, T.~P.}
\newblock \bibinfo{title}{Low-temperature kinetics and dynamics with {C}oulomb
  crystals}.
\newblock \emph{\bibinfo{journal}{Annu. Rev. Phys. Chem.}}
  \textbf{\bibinfo{volume}{66}}, \bibinfo{pages}{475--495}
  (\bibinfo{year}{2015}).

\bibitem{willitsch17a}
\bibinfo{author}{Willitsch, S.}
\newblock \bibinfo{title}{Chemistry with controlled ions}.
\newblock \emph{\bibinfo{journal}{Adv. Chem. Phys.}}
  \textbf{\bibinfo{volume}{162}}, \bibinfo{pages}{307--340}
  (\bibinfo{year}{2017}).

\bibitem{filsinger08a}
\bibinfo{author}{Filsinger, F.}, \bibinfo{author}{Erlekam, U.},
  \bibinfo{author}{von Helden, G.}, \bibinfo{author}{K\"upper, J.} \&
  \bibinfo{author}{Meijer, G.}
\newblock \bibinfo{title}{Selector for structural isomers of neutral
  molecules}.
\newblock \emph{\bibinfo{journal}{Phys. Rev. Lett.}}
  \textbf{\bibinfo{volume}{100}}, \bibinfo{pages}{133003}
  (\bibinfo{year}{2008}).

\bibitem{filsinger09a}
\bibinfo{author}{Filsinger, F.} \emph{et~al.}
\newblock \bibinfo{title}{Pure samples of individual conformers: the separation
  of stereo-isomers of complex molecules using electric fields}.
\newblock \emph{\bibinfo{journal}{Angew. Chem. Int. Ed.}}
  \textbf{\bibinfo{volume}{48}}, \bibinfo{pages}{6900} (\bibinfo{year}{2009}).

\bibitem{horke14a}
\bibinfo{author}{Horke, D.~A.}, \bibinfo{author}{Chang, Y.-P.},
  \bibinfo{author}{D{\l}ugo{\l}{\k{e}}cki, K.} \& \bibinfo{author}{K\"upper,
  J.}
\newblock \bibinfo{title}{Separating para and ortho water}.
\newblock \emph{\bibinfo{journal}{Angew. Chem. Int. Ed.}}
  \textbf{\bibinfo{volume}{53}}, \bibinfo{pages}{11965--11968}
  (\bibinfo{year}{2014}).

\bibitem{trippel18a}
\bibinfo{author}{Trippel, S.} \emph{et~al.}
\newblock \bibinfo{title}{Knife edge skimming for improved separation of
  molecular species by the deflector}.
\newblock \emph{\bibinfo{journal}{Rev. Sci. Instrum.}}
  \textbf{\bibinfo{volume}{89}}, \bibinfo{pages}{096110}
  (\bibinfo{year}{2018}).

\bibitem{chang13a}
\bibinfo{author}{Chang, Y.-P.} \emph{et~al.}
\newblock \bibinfo{title}{Specific chemical reactivities of spatially separated
  3-aminophenol conformers with cold {Ca$^{+}$} ions}.
\newblock \emph{\bibinfo{journal}{Science}} \textbf{\bibinfo{volume}{342}},
  \bibinfo{pages}{98} (\bibinfo{year}{2013}).

\bibitem{roesch14a}
\bibinfo{author}{R{\"o}sch, D.}, \bibinfo{author}{Willitsch, S.},
  \bibinfo{author}{Chang, Y.-P.} \& \bibinfo{author}{K{\"u}pper, J.}
\newblock \bibinfo{title}{Chemical reactions of conformationally selected
  3-aminophenol molecules in a beam with coulomb-crystallized {Ca$^{+}$} ions}.
\newblock \emph{\bibinfo{journal}{J. Chem. Phys.}}
  \textbf{\bibinfo{volume}{140}}, \bibinfo{pages}{124202}
  (\bibinfo{year}{2014}).

\bibitem{kilaj18a}
\bibinfo{author}{Kilaj, A.} \emph{et~al.}
\newblock \bibinfo{title}{Observation of different reactivities of para and
  ortho-water towards trapped diazenylium ions}.
\newblock \emph{\bibinfo{journal}{Nat. Commun.}} \textbf{\bibinfo{volume}{9}},
  \bibinfo{pages}{2096} (\bibinfo{year}{2018}).

\bibitem{kilaj20a}
\bibinfo{author}{Kilaj, A.} \emph{et~al.}
\newblock \bibinfo{title}{Quantum-chemistry-aided identification, synthesis and
  experimental validation of model systems for conformationally controlled
  reaction studies: Separation of the conformers of 2,3-dibromobuta-1,3-diene
  in the gas phase}.
\newblock \emph{\bibinfo{journal}{Phys. Chem. Chem. Phys.}}
  \textbf{\bibinfo{volume}{22}}, \bibinfo{pages}{13431--13439}
  (\bibinfo{year}{2020}).

\bibitem{roesch16a}
\bibinfo{author}{R\"osch, D.}, \bibinfo{author}{Gao, H.},
  \bibinfo{author}{Kilaj, A.} \& \bibinfo{author}{Willitsch, S.}
\newblock \bibinfo{title}{Design and characterization of a linear quadrupole
  ion trap for high-resolution coulomb-crystal time-of-flight mass
  spectrometry}.
\newblock \emph{\bibinfo{journal}{EPJ Tech. Instrum.}}
  \textbf{\bibinfo{volume}{3}}, \bibinfo{pages}{5} (\bibinfo{year}{2016}).

\bibitem{clary87a}
\bibinfo{author}{Clary, D.}
\newblock \bibinfo{title}{Rate constants for the reactions of ions with dipolar
  polyatomic molecules}.
\newblock \emph{\bibinfo{journal}{J. Chem. Soc., Faraday Trans. 2}}
  \textbf{\bibinfo{volume}{83}}, \bibinfo{pages}{139--148}
  (\bibinfo{year}{1987}).

\bibitem{stoecklin92a}
\bibinfo{author}{Stoecklin, T.}, \bibinfo{author}{Clary, D.~C.} \&
  \bibinfo{author}{Palma, A.}
\newblock \bibinfo{title}{Rate constant calculations for ion-symmetric top and
  ion-asymmetric top reactions}.
\newblock \emph{\bibinfo{journal}{J. Chem. Soc. Faraday Trans.}}
  \textbf{\bibinfo{volume}{88}}, \bibinfo{pages}{901} (\bibinfo{year}{1992}).

\bibitem{zhao08b}
\bibinfo{author}{Zhao, Y.} \& \bibinfo{author}{Truhlar, D.~G.}
\newblock \bibinfo{title}{The {M06} suite of density functionals for main group
  thermochemistry, thermochemical kinetics, noncovalent interactions, excited
  states, and transition elements: two new functionals and systematic testing
  of four {M06}-class functionals and 12 other functionals}.
\newblock \emph{\bibinfo{journal}{Theor. Chem. Acc.}}
  \textbf{\bibinfo{volume}{120}}, \bibinfo{pages}{215} (\bibinfo{year}{2008}).

\bibitem{linder13a}
\bibinfo{author}{Linder, M.} \& \bibinfo{author}{Brinck, T.}
\newblock \bibinfo{title}{On the method-dependence of transition state
  asynchonicity in {D}iels-{A}lder reactions}.
\newblock \emph{\bibinfo{journal}{Phys. Chem. Chem. Phys.}}
  \textbf{\bibinfo{volume}{15}}, \bibinfo{pages}{5108} (\bibinfo{year}{2013}).

\bibitem{weigend05a}
\bibinfo{author}{Weigend, F.} \& \bibinfo{author}{Ahlrichs, R.}
\newblock \bibinfo{title}{Balanced basis sets of split valence, triple zeta
  valence and quadruple zeta valence quality for {H} to {R}n: {D}esign and
  assessment of accuracy}.
\newblock \emph{\bibinfo{journal}{Phys. Chem. Chem. Phys.}}
  \textbf{\bibinfo{volume}{7}}, \bibinfo{pages}{3297--3305}
  (\bibinfo{year}{2005}).

\bibitem{g09}
\bibinfo{author}{Frisch, M.~J.} \emph{et~al.}
\newblock \bibinfo{title}{Gaussian 09 {R}evision {D}.01}.
\newblock \bibinfo{note}{{G}aussian Inc., Wallingford CT, 2009}.

\bibitem{bergner1993ab}
\bibinfo{author}{Bergner, A.}, \bibinfo{author}{Dolg, M.},
  \bibinfo{author}{K{\"u}chle, W.}, \bibinfo{author}{Stoll, H.} \&
  \bibinfo{author}{Preu{\ss}, H.}
\newblock \bibinfo{title}{Ab initio energy-adjusted pseudopotentials for
  elements of groups 13--17}.
\newblock \emph{\bibinfo{journal}{Mol. Phys.}} \textbf{\bibinfo{volume}{80}},
  \bibinfo{pages}{1431--1441} (\bibinfo{year}{1993}).

\bibitem{lee88b}
\bibinfo{author}{Lee, C.}, \bibinfo{author}{Yang, W.} \& \bibinfo{author}{Parr,
  R.~G.}
\newblock \bibinfo{title}{Development of the {C}olle-{S}alvetti
  correlation-energy formula into a functional of the electron density}.
\newblock \emph{\bibinfo{journal}{Phys. Rev. B}} \textbf{\bibinfo{volume}{37}},
  \bibinfo{pages}{785--789} (\bibinfo{year}{1988}).

\bibitem{becke93a}
\bibinfo{author}{Becke, A.~D.}
\newblock \bibinfo{title}{Density functional thermochemistry. {III}. {T}he role
  of exact exchange}.
\newblock \emph{\bibinfo{journal}{J. Chem. Phys.}}
  \textbf{\bibinfo{volume}{98}}, \bibinfo{pages}{5648--5652}
  (\bibinfo{year}{1993}).

\bibitem{knizia2013}
\bibinfo{author}{Knizia, G.}
\newblock \bibinfo{title}{Intrinsic atomic orbitals: An unbiased bridge between
  quantum theory and chemical concepts}.
\newblock \emph{\bibinfo{journal}{J. Chem. Theory Comput.}}
  \textbf{\bibinfo{volume}{9}}, \bibinfo{pages}{4834--4843}
  (\bibinfo{year}{2013}).

\bibitem{schmid17a}
\bibinfo{author}{Schmid, P.~C.}, \bibinfo{author}{Greenberg, J.},
  \bibinfo{author}{Miller, M.~I.}, \bibinfo{author}{Loeffler, K.} \&
  \bibinfo{author}{Lewandowski, H.~J.}
\newblock \bibinfo{title}{An ion trap time-of-flight mass spectrometer with
  high mass resolution for cold trapped ion experiments}.
\newblock \emph{\bibinfo{journal}{Rev. Sci. Instrum}}
  \textbf{\bibinfo{volume}{88}}, \bibinfo{pages}{123107}
  (\bibinfo{year}{2017}).

\bibitem{hankin01a}
\bibinfo{author}{Hankin, S.~M.}, \bibinfo{author}{Villeneuve, D.~M.},
  \bibinfo{author}{Corkum, P.~B.} \& \bibinfo{author}{Rayner, D.~M.}
\newblock \bibinfo{title}{Intense-field laser ionization rates in atoms and
  molecules}.
\newblock \emph{\bibinfo{journal}{Phys. Rev. A}} \textbf{\bibinfo{volume}{64}},
  \bibinfo{pages}{013405} (\bibinfo{year}{2001}).

\bibitem{wiese19a}
\bibinfo{author}{Wiese, J.}, \bibinfo{author}{Olivieri, J.-F.},
  \bibinfo{author}{Trabattoni, A.}, \bibinfo{author}{Trippel, S.} \&
  \bibinfo{author}{K\"upper, J.}
\newblock \bibinfo{title}{Strong-field photoelectron momentum imaging of {OCS}
  at finely resolved incident intensities}.
\newblock \emph{\bibinfo{journal}{New J. Phys}} \textbf{\bibinfo{volume}{21}},
  \bibinfo{pages}{083011} (\bibinfo{year}{2019}).

\bibitem{simion11a}
\bibinfo{author}{Manura, D.} \& \bibinfo{author}{Dahl, D.}
\newblock \emph{\bibinfo{title}{{SIMION 8.0/8.1 User manual}}}.
\newblock \bibinfo{organization}{Scientific Instrument Services, Inc.},
  \bibinfo{address}{Ringoes, NJ}, \bibinfo{edition}{rev. 5} edn.
  (\bibinfo{year}{2011}).

\bibitem{goebbert04a}
\bibinfo{author}{Goebbert, D.~J.}, \bibinfo{author}{Liu, X.} \&
  \bibinfo{author}{Wenthold, P.~G.}
\newblock \bibinfo{title}{Reactions of diacetylene radical cation with
  ethylene}.
\newblock \emph{\bibinfo{journal}{J. Am. Soc. Mass Spectrom.}}
  \textbf{\bibinfo{volume}{15}}, \bibinfo{pages}{114--120}
  (\bibinfo{year}{2004}).

\bibitem{colorado98a}
\bibinfo{author}{Colorado, A.}, \bibinfo{author}{Barket, J.~D.},
  \bibinfo{author}{Hurst, M.~J.} \& \bibinfo{author}{Shepson, B.~P.}
\newblock \bibinfo{title}{A fast-response method for determination of
  atmospheric isoprene using quadrupole ion trap mass spectrometry}.
\newblock \emph{\bibinfo{journal}{Anal. Chem.}} \textbf{\bibinfo{volume}{70}},
  \bibinfo{pages}{5129--5135} (\bibinfo{year}{1998}).

\bibitem{hofmann99b}
\bibinfo{author}{Hofmann, M.} \& \bibinfo{author}{Schaefer, H.~F.}
\newblock \bibinfo{title}{Pathways for the reaction of the butadiene radical
  cation {C}$_{4}${H}$_{6}^{+}$, with ethylene}.
\newblock \emph{\bibinfo{journal}{J. Phys. Chem. A}}
  \textbf{\bibinfo{volume}{103}}, \bibinfo{pages}{8895--8905}
  (\bibinfo{year}{1999}).

\bibitem{bouchoux94a}
\bibinfo{author}{Bouchoux, G.}, \bibinfo{author}{Salpin, J.-Y.} \&
  \bibinfo{author}{Turecek, F.}
\newblock \bibinfo{title}{Cycloaddition reactions between 1,3-butadiene radical
  cations and ethene in the gas phase}.
\newblock \emph{\bibinfo{journal}{Rapid Commun. Mass Spectrom}}
  \textbf{\bibinfo{volume}{8}}, \bibinfo{pages}{325--328}
  (\bibinfo{year}{1994}).

\bibitem{gross17a}
\bibinfo{author}{Gross, J.}
\newblock \emph{\bibinfo{title}{Mass Spectrometry}}
  (\bibinfo{publisher}{Springer International Publishing},
  \bibinfo{address}{Cham}, \bibinfo{year}{2017}), \bibinfo{edition}{3} edn.

\bibitem{kuck85a}
\bibinfo{author}{Kuck, D.}, \bibinfo{author}{Schneider, J.} \&
  \bibinfo{author}{Grützmacher, H.-F.}
\newblock \bibinfo{title}{A study of gaseous benzenium and toluenium ions
  generated from 1,4- dihydro- and 1-methyl-1,4-dihydro-benzoic acids}.
\newblock \emph{\bibinfo{journal}{J. Chem. Soc., Perkin Trans. 2}}
  \bibinfo{pages}{689--696} (\bibinfo{year}{1985}).

\bibitem{schroeder06a}
\bibinfo{author}{Schröder, D.}, \bibinfo{author}{Schwarz, H.},
  \bibinfo{author}{Milko, P.} \& \bibinfo{author}{Roithov{\'{a}}, J.}
\newblock \bibinfo{title}{Dissociation routes of protonated toluene probed by
  infrared spectroscopy in the gas phase}.
\newblock \emph{\bibinfo{journal}{J. Phys. Chem. A}}
  \textbf{\bibinfo{volume}{110}}, \bibinfo{pages}{8346--8353}
  (\bibinfo{year}{2006}).

\bibitem{wang18a}
\bibinfo{author}{Wang, Z.-C.} \emph{et~al.}
\newblock \bibinfo{title}{The gas-phase methylation of benzene and toluene}.
\newblock \emph{\bibinfo{journal}{Int. J. Mass Spectrom.}}
  \textbf{\bibinfo{volume}{429}}, \bibinfo{pages}{6--13}
  (\bibinfo{year}{2018}).

\bibitem{lifshitz94a}
\bibinfo{author}{Lifshitz, C.}
\newblock \bibinfo{title}{Tropylium ion formation from toluene: Solution of an
  old problem in organic mass spectrometry}.
\newblock \emph{\bibinfo{journal}{Acc. Chem. Res.}}
  \textbf{\bibinfo{volume}{27}}, \bibinfo{pages}{138--144}
  (\bibinfo{year}{1994}).

\end{thebibliography}


\clearpage
\setcounter{section}{0}
\setcounter{figure}{0}
\setcounter{table}{0}
\setcounter{equation}{0}

\renewcommand\thesection{S\arabic{section}}
\renewcommand\thefigure{S\arabic{figure}}
\renewcommand\thetable{S\arabic{table}}
\renewcommand\theequation{S\arabic{equation}}

\section*{Supplementary Information}

\section{Molecular beam density}

The density of DBB molecules in the molecular beam, $n_\mathrm{DBB}$, was determined in order to convert the measured pseudo-first-order rate constants into bimolecular rate constants. For this purpose, the sensitivity of the TOF-MS was first calibrated to be able to convert the integrated MCP signal into an absolute number of ions \cite{schmid17a}. Using this information, the DBB density was determined from the dependence of the photoion yield on laser intensity in a strong-field multi-photon ionisation experiment \cite{hankin01a, wiese19a}. 

\subsection{Calibration of the TOF-MS sensitivity}

\begin{figure}[h]
\centering
\includegraphics[width=80 mm]{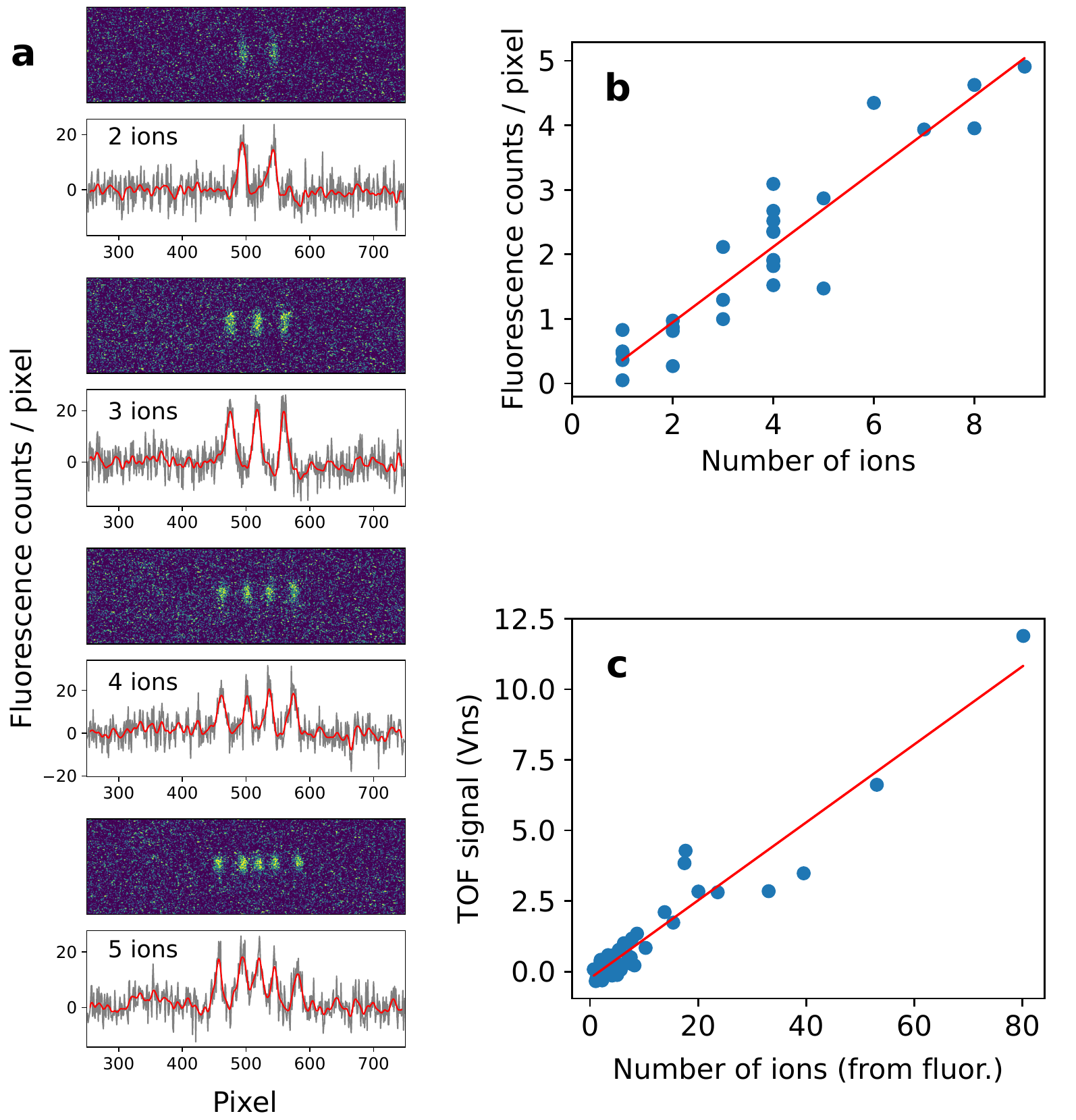}
\caption{\textbf{Calibration of the TOF-MS.} \textbf{a}, Fluorescence images of Ca$^{+}$ Coulomb crystals with 2 to 5 ions (top to bottom). The fluorescence integrated along the vertical image axis is shown below each image. The red line corresponds to a moving average of over 11 points. \textbf{b} Integrated Ca$^{+}$ fluorescence counts as a function ion number. The red line corresponds to a linear function fitted to the data. \textbf{c}~Integrated Ca$^{+}$ TOF-MS signal as a function of ion number (determined from measured fluorescence counts based on the fit function in \textbf{b}. The red line corresponds to a linear function fitted to the data.}
\label{si_fig_tofms_calib}
\end{figure}

In order to calibrate the sensitivity of the TOF-MS, the TOF-MS signals generated by Coulomb crystals of a well-defined ion number were measured \cite{schmid17a}. Fig.~\ref{si_fig_tofms_calib}a shows exemplary fluorescence images of Ca$^{+}$ ion strings containing two to five ions. Counting the number of ions in each image was facilitated by integrating the fluorescence in the images along the vertical axis. The corresponding integrated fluorescence curves are shown below the images with the red line corresponding to a moving average over 11 points. Fig.~\ref{si_fig_tofms_calib}b shows the correlation between the number of ions and the total integrated fluorescence counts for a selection of 28 Coulomb crystal images with up to 9 ions. Fitting the data with a linear function gives a slope of $0.65(6)$ counts per ion. This relation enabled the determination of the number of ions using the integrated fluorescence of any Ca$^{+}$ Coulomb crystal. For a larger set of Coulomb crystals, both fluorescence images as well as TOF-MS data were acquired. Fig.~\ref{si_fig_tofms_calib}c shows the integrated Ca$^{+}$ TOF-MS signal as a function of ion number, determined from fluorescence counts. Fitting these data with a linear function gives a TOF-MS sensitivity of $0.136(14)~\mathrm{V ns/ion}$, where the uncertainty also includes the uncertainty in the relation between ion number and fluorescence counts.

\subsection{Determination of the density of the DBB molecular beam}

\begin{figure}[h]
\centering
\includegraphics[width=80 mm]{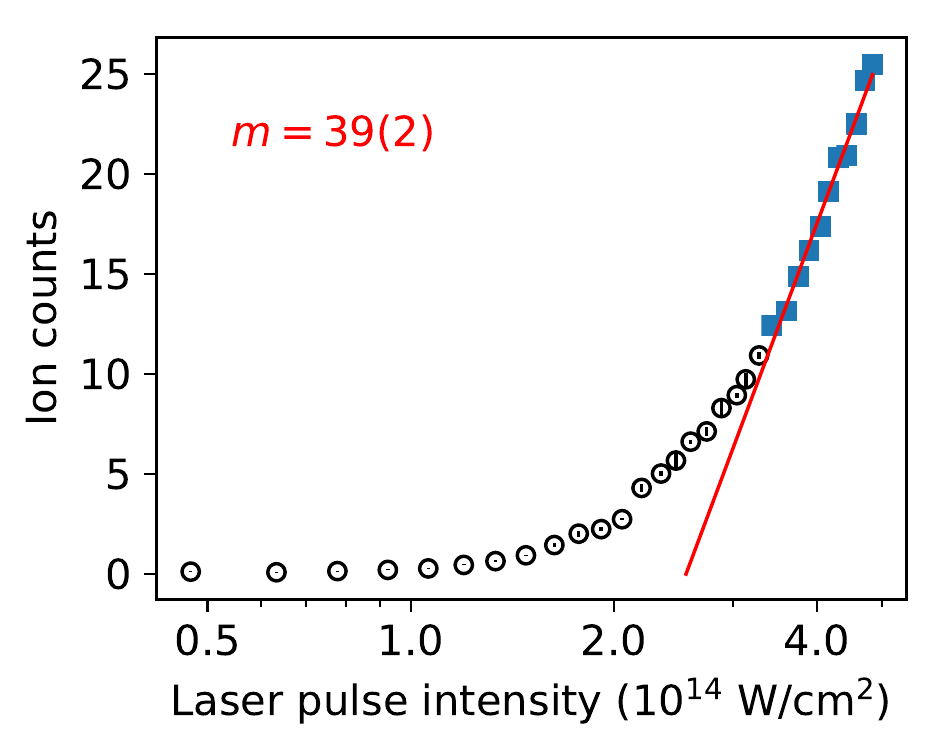}
\caption{\textbf{Calibration of the DBB beam density.} The ion yield from fs-laser photoionisation was measured as a function of the laser intensity. At intensities $I>3.3\times10^{14}~\mathrm{W/cm}^{2}$ (blue squares), the ion yield is well represented by a logarithmic dependence on the laser intensity (red line) from which the beam density can be determined (see text).}
\label{si_fig_beam_density}
\end{figure}

The density of DBB molecules in the molecular beam was determined using ionisation yields in non-resonant strong-field ionisation \cite{hankin01a, wiese19a}. Here, a voltage of 13~kV was applied to the deflector and the molecular-beam apparatus was set to a deflection coordinate of $y=0~$mm. The molecular beam was ionised by multi-photon ionisation using laser pulses with a duration of  150~fs at 775~nm. The total ion yield was measured as a function of the laser-pulse intensity,  Fig.~\ref{si_fig_beam_density} and was found to scale logarithmically with laser intensity $I$ for $I>3.3\times 10^{14}~\mathrm{W/cm^{2}}$. This behaviour is expected for saturated multi-photon ionisation \cite{hankin01a} and allows the extraction of the molecular-beam peak density $n_\mathrm{peak}$ from the slope of the line \cite{wiese19a}
\begin{equation}
m := \frac{\mathrm{d}S}{\mathrm{d}\ln{I/I_0}} = 2\pi \alpha \sigma_r^2 d n_\mathrm{peak}
\label{si_eq_beam_density}
\end{equation}
where $S$ is the ion count, $I$ is the laser intensity, $I_0$ is the saturation intensity of the relevant molecular species, $\alpha = 0.57$ is the detection efficiency, $\sigma_r = 7.5~\mu$m is the $e^{-1/2}$ radius of the laser beam and $d = 2.5$~mm is the diameter of the molecular beam. We determined $m = 39(2)$ and $I_0 = 2.56(5)\times 10^{14}~\mathrm{W/cm^{2}}$ from a linear fit of $S$ as a function of $\ln(I)$ (Fig.~\ref{si_fig_beam_density}). Using  \eqref{si_eq_beam_density}, a DBB density of $n_\mathrm{peak} = 7.8(7)\times 10^{7}~\mathrm{cm^{-3}}$ was obtained. The uncertainty includes the error of the TOF-MS calibration. Because the molecular beam was pulsed with a repetition rate $f_\mathrm{rep} = 200$~Hz and a pulse duration of $\tau_\mathrm{pulse} = 250~\mu$s at the position of the LQT, the time-averaged density relevant for the reaction experiments was calculated to be $n_\mathrm{avg} = n_\mathrm{peak} f_\mathrm{rep}\tau_\mathrm{pulse} = 3.9(4)\times10^6~\mathrm{cm^{-3}}$.

\section{TOF mass spectra of reaction products}

\subsection{Molecular dynamics simulations}

\begin{figure}[h!]
\centering
\includegraphics[width=120 mm]{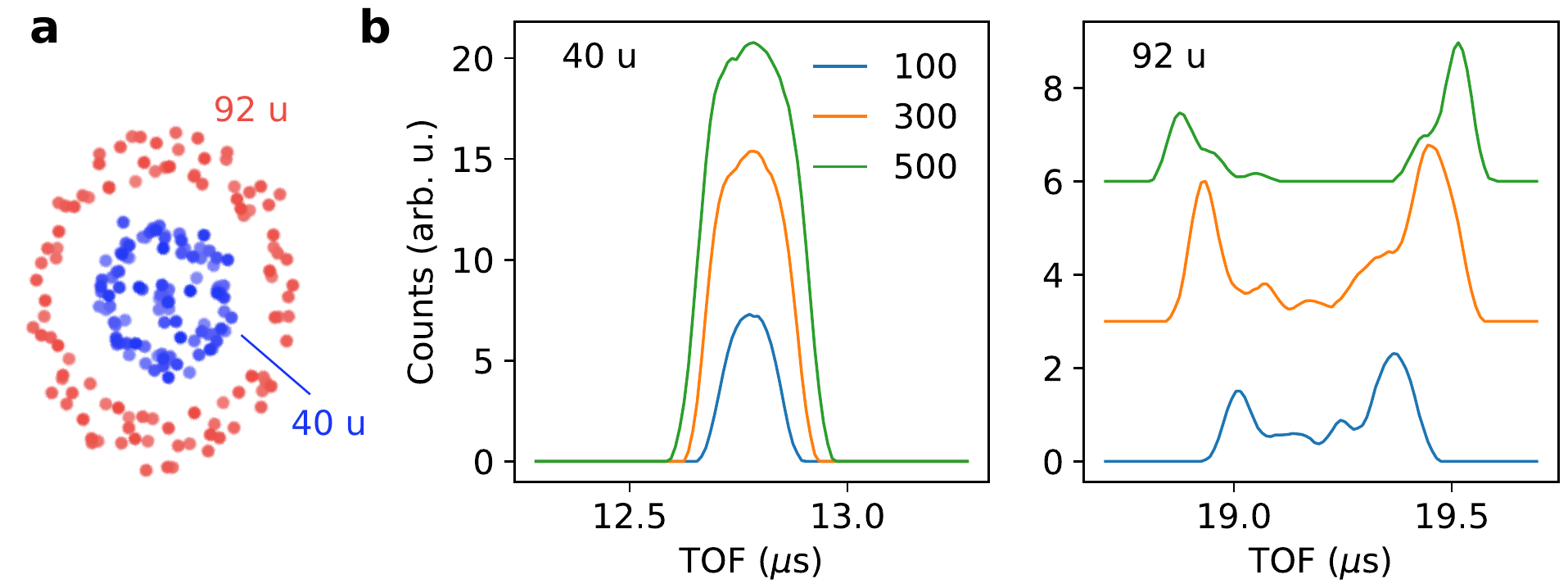}
\caption{\textbf{Influence of Coulomb-crystal size on ion time-of-flight spectra}. \textbf{a} Simulated Coulomb crystal consisting of ions with mass of 40~u (blue) and mass 92~u (red). \textbf{b} Simulated TOF mass spectra for Coulomb crystals consisting of a variable number (100, 300, 500) of ions with mass 40~u and 100 ions of mass 92~u. With increasing number of lighter ions, the width of the bimodal time-of-flight distribution for the ions with mass 92~u increases. TOF traces for the species at mass 92~u are vertically offset for clarity.}
\label{si_fig_simion_92}
\end{figure}

For an accurate assignment of the signals in the TOF mass spectra shown in Fig.~\ref{fig_products_tofms} of the main text, molecular dynamics (MD) simulations of mixed species Coulomb crystals ejected into the TOF-MS were performed using the SIMION software \cite{simion11a}. In an ion Coulomb crystal, ions spatially arrange in shells according to their mass-to-charge ratio. While lighter ions accumulate closer to the centre, heavier ions localise around the lighter particles leading to an onion-like arrangement of the different ion species in the crystal \cite{willitsch12a} (Fig.~\ref{si_fig_simion_92}a). In the present experiments, the central core of the crystals is formed by the laser-cooled Ca$^{+}$ ions. 

The shell-like ion arrangements strongly affect the ion dynamics during ejection into the TOF-MS flight tube and lead to a dispersion of the time-of-flight of the ions depending on their initial position in the crystal. The extended ion packets impinging on the detector give rise to bimodal distributions in the TOF spectrum with maxima at two distinct arrival times \cite{roesch16a}. The widths of the bimodal TOF signals strongly depend on the amount of lighter ions in the original Coulomb crystal which determine the diameter of the shells of heavier ions.

To illustrate this effect, MD simulations of Coulomb crystals were performed which are composed of two ion species with mass 40~u and 92~u. The number of ions with mass 92~u was fixed to 100 and the number of ions of mass 40~u was varied between 100 and 500. The electrode potentials in the simulations were chosen to be identical to the ones in the experiment. Simulated TOF mass spectra of these crystals are presented in Fig.~\ref{si_fig_simion_92}b. While the lighter species produces a single peak in the TOF trace, the heavier ions show a bimodal distribution the width of which increases with the number of light ions.

\begin{figure}[!h]
\centering
\includegraphics[width=120 mm]{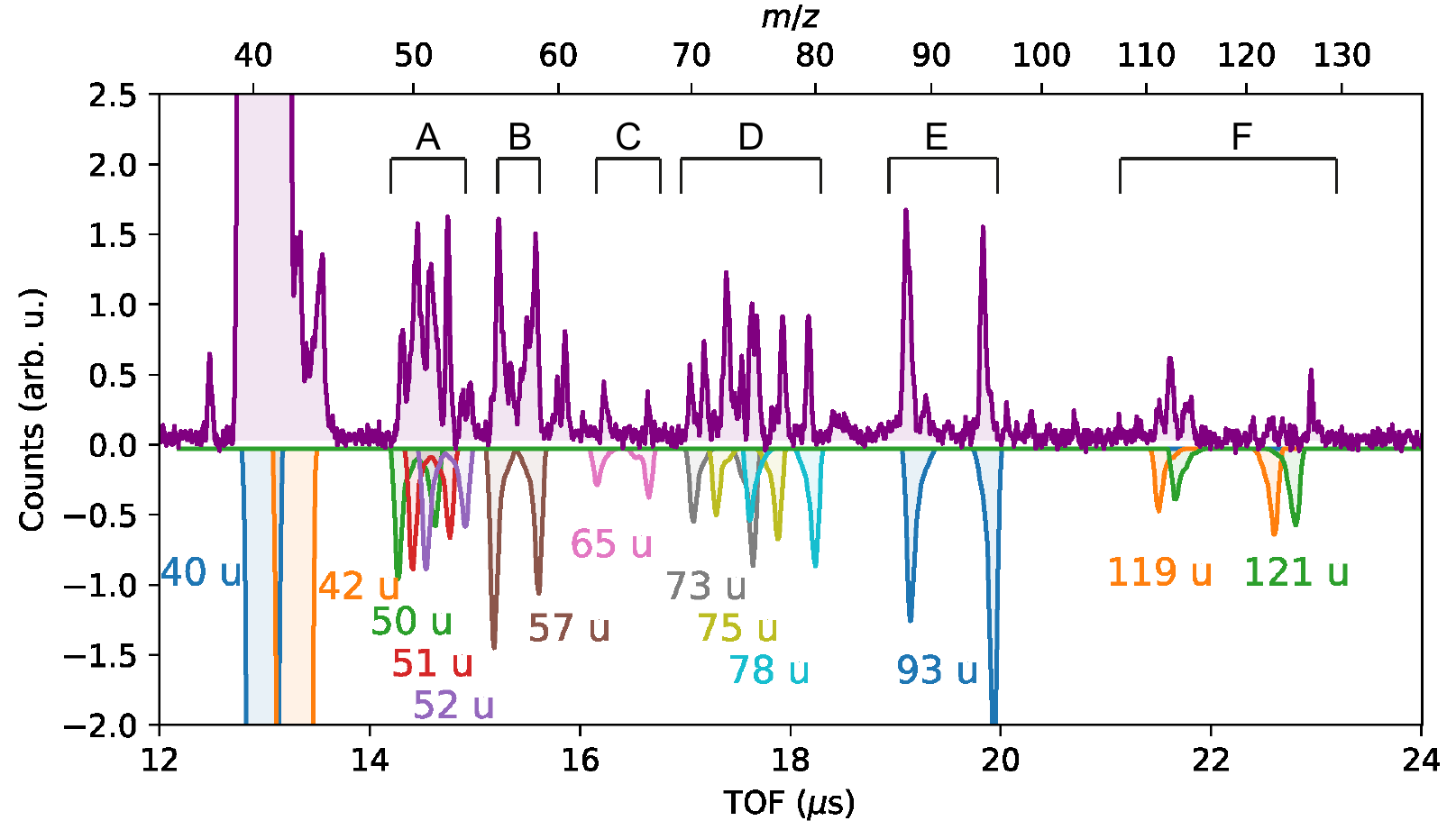}
\caption{\textbf{Experimental and simulated TOF mass spectra.} The experimental TOF mass spectrum of Fig. \ref{fig_products_tofms} in the main text (top trace) is compared with a simulation (lower inverted trace). Features corresponding to different masses are highlighted in different colours for clarity. The estimated uncertainty of the mass determination is $\pm$1~u, see text for discussion.}
\label{si_fig_comparison_theory}
\end{figure}

In order to assign masses to the different bands observed in the product TOF-MS spectrum of Fig.~\ref{fig_products_tofms} of the main text, MD simulations of Coulomb crystals consisting of Ca$^{+}$ ions (mass 40~u) and different compositions of molecular ions corresponding to the reactants and products were performed. A comparison of the experimental data with a best-fit simulation is shown in Fig.~\ref{si_fig_comparison_theory}. The corresponding ion composition of the Coulomb crystal assumed in this simulation is detailed in Tab. \ref{tab_sim}. For clarity, the simulated TOF signals are shown in different colours to highlight the contribution from each mass. A global time offset of $\lesssim 0.5~\mu s$ was added to the simulated spectrum in order to match the position of the Ca$^{+}$ peak in the experimental data. We find an overall satisfactory agreement between experiment and simulation. In particular, the observed splittings of the individual mass signals are reproduced in the simulated TOF spectra. Small differences in the splittings and positions of the peaks in the simulation are attributed to different sizes and compositions of the Coulomb crystal in the experiment and simulations and to the possible presence of additional heavy ion species which are not accounted for in the simulations. Because of these uncertainties, we estimate the accuracy of the determination of the masses to be $\pm$1~u.

\begin{table}[h]
\begin{center}
\begin{tabular}{c|c||c|c}
Ion mass (u) & Number of ions & Ion mass (u) & Number of ions \\ \hline
40 & 450 & 73 & 20 \\
42 & 210 & 75 & 16\\
50 & 20 & 78 & 20 \\
51 & 20 & 93 & 48 \\
52 & 20 & 119 & 16 \\
57 & 36 & 121 & 16 \\
65 & 10 & & \\
\end{tabular}
\caption{\textbf{Ion composition of Coulomb crystal} for the simulated TOF mass spectrum of Fig. \ref{si_fig_comparison_theory}.}
\label{tab_sim}
\end{center}
\end{table}

Based on the agreement between experiment and simulation, we assign the bands A--F in Fig.~~\ref{fig_products_tofms} of the main text to different molecular formulae as summarised in table~\ref{si_tab_masses}.

\begin{table}[h]
\begin{center}
\begin{tabular}{c | c | l}
Label & Assigned masses (u) & Possible compounds\\\hline
A 	& 50,51,52	& $\mathrm{C}_4\mathrm{H}_n^{+}$, $n$=2,3,4\\
B	& 57 & $\mathrm{CaOH^{+}}$\\
C	& 65 & $\mathrm{C_5H_5^{+}}$\\
D	& 77,78 & $\mathrm{C}_6\mathrm{H}_n^{+}$, $n=5,6$\\
E	& 93 & $\mathrm{C_7H_9^+}$\\
F	& 119, 121 & CaBr$^{+}$
\end{tabular}
\caption{\textbf{Assignment of the main TOF bands originating from the reaction of DBB with propene ions} (not accounting for masses produced by background reactions). The estimated uncertainty of the mass determination is $\pm$1~u, see text for discussion. }
\label{si_tab_masses}
\end{center}

\end{table}

\subsection{Product fragmentation pathways}

The fragmentation of unstable products of radical cation reactions in the gas phase is a widely observed phenomenon, see, e.g., Refs. \cite{eberlin04a, goebbert04a, colorado98a, hofmann99b,  bouchoux94a, gross17a}. 
Considering the excess energy of $>60$~kcal/mol of the present reaction and the constant presence of near-infrared and near-ultraviolet laser light in the experiment, a range of different pathways leading from the DA cycloadduct to the observed fragments listed in Tab. \ref{si_tab_masses} are conceivable. It is plausible that in a first step, a cyclic C$_7$H$_{10}^+$ moiety ($m/z=94$~u) is formed through the loss of the two bromine atoms from the DA cycloadduct, the 2-dibromo-4-methyl-cyclohexene radical cation (Fig. 1a). This fragment ion can further undergo hydrogen loss forming protonated toluene (93~u, band E in Fig. \ref{fig_products_tofms} of the main text). Elimination of a methyl group or methane can result in the benzene (78~u) and further H loss in the phenyl (77~u) ion  (band D) \cite{kuck85a, schroeder06a, wang18a}. As an alternative, a sequence of hydrogen losses, possibly involving a skeletal rearrangement to the tropylium ion \cite{lifshitz94a}, and the elimination of C$_2$H$_2$ results in the formation of the cyclopentadienyl cation C$_5$H$_5^+$ (65~u, band C)\cite{gross17a}. The C$_4$H$_n^+$ fragments (band A) could partially result from further breakup of this C$_5$ moiety, but are mostly accounted for by background reactions of DBB with Ca$^+$ (see main text).

In order to achieve a more detailed understanding of the fragmentation pathways, further studies at a higher mass resolution which would enable the unambiguous determination of the fragment masses in combination with theoretical calculations of the potential energy surfaces would be required. However, we note the striking general resemblance of the fragmentation pattern of the product of the reaction of DBB with propene ions shown in Fig. \ref{si_fig_comparison_theory} (and Fig. \ref{fig_products_tofms} of the main text) with the ones observed in the mass spectra of cyclic compounds such as toluene \cite{gross17a}. This similarity provides further evidence that the product of the title reaction is indeed the cyclic DA adduct.

\section{Potential energy surface}

\begin{table}[h!]
\centering
\caption{\textbf{Single-point energies (SPE) of stationary points on the PES of the reaction system and $\langle S^2\rangle$ values} calculated at the M06-2X/def2-TZVPP level of theory with Stuttgart ECP including relativistic effects on the bromine atoms \cite{bergner1993ab}. Refer to Fig. \ref{fig_theory} of the main text for the nomenclature of the structures.}
\begin{tabular}{ c | c | c } 
Structure & \begin{tabular}{@{}c@{}}SPE \\ (kcal/mol)\end{tabular}  & $\langle{S^{2}}\rangle$\\\hline
gauche & 0.00 & -\\
s-trans & -0.50 & -\\
I1$^{g}$ & -12.40 & 0.756\\
I1$^{t}$ & -12.97 & 0.756\\
I2$^{g}$ & -20.18 & 0.783\\
I2$^{t}$ & -25.16 & 0.790\\
I3$^{g}$ & -10.72 & 0.757\\
P1$^{g}$ & -58.75 & 0.753\\
P1$^{t}$ & -31.72 & 0.777\\
P2$^{g}$ & -62.40 & 0.753\\
P2$^{t}$ & -33.29 & 0.778\\
P3$^{g}$ & -31.90 & 0.780\\
P3$^{t}$ & -32.24 & 0.780\\
TS1$^{g}$ & -9.91 & 0.758\\
TS1$^{t}$ & -12.40 & 0.757\\
TS2$^{g}$ & -17.91 & 0.781\\
TS2$^{t}$ & -22.49 & 0.789\\
TS3$^{g}$ & -10.11 & 0.756\\
TS4$^{g}$ & -18.34 & 0.786\\
TS4$^{t}$ & -23.18 & 0.790\\
TS5$^{g}$ & -14.11 & 0.784\\
TS5$^{t}$ & -18.52 & 0.790\\
TS6$^{g}$ & -58.55 & 0.753\\
TS6$^{t}$ & -31.61 & 0.777\\
TS7 & -16.30 & 0.7752\\
TS8 & -24.48 & 0.7596\\
\end{tabular}
\label{si_tab_SPE_S2}
\end{table}

\end{document}